\documentclass[pre,twocolumn,twoside,byrevtex,superscriptaddress,floatfix]{revtex4-1}

\usepackage{morefloats}

\usepackage{url}
\usepackage{graphicx,epsfig,verbatim,enumerate}
\usepackage{amssymb,amsmath}
\usepackage{ifthen}
\newboolean{twocolswitch}

\usepackage[table]{xcolor}

\definecolor{lightgrey}{rgb}{0.5,0.5,0.5}
\definecolor{verylightgrey}{rgb}{0.9,0.9,0.9}

\usepackage{rotating}
\usepackage{longtable}
\usepackage{array}

\newcommand{\sindex}[1]{}
\newcommand{\nindex}[1]{}

\newcommand{\www}[1]{\url{#1}}

\newcommand{\PreserveBackslash}[1]{\let\temp=\\#1\let\\=\temp}
\newcommand{\PBS}[1]{\let\temp=\\#1\let\\=\temp}

\newcommand{\havg}[1]{h_{\rm avg}(#1)}
\newcommand{\havgword}[1]{h_{\rm avg}(\textnormal{`#1'})}

\newcommand{\havgfn}{h_{\rm avg}}
\newcommand{\hstdfn}{h_{\textnormal{std}}}

\newcommand{\Probof}[1]{\mathbf{Pr}(#1)}

\newcommand{\Tref}{T_{\textnormal{ref}}}
\newcommand{\Tcomp}{T_{\textnormal{comp}}}

\newcommand{\deltahw}{\delta_{h}(w)}
\newcommand{\deltapw}{\delta_{p}(w)}

\usepackage[framemethod=default]{mdframed}
\newmdenv[linecolor=lightgrey,backgroundcolor=white]{myframe}

\setboolean{twocolswitch}{true}

\newcommand{\revtexonly}[1]{#1}
\newcommand{\plainlatexonly}[1]{}

\begin{document}

\title{Human language reveals a universal positivity bias
}

\author{
\firstname{Peter Sheridan}
\surname{Dodds}
}
\email{peter.dodds@uvm.edu}

\affiliation{
  Computational Story Lab,
  Vermont Advanced Computing Core,
  \& the Department of Mathematics and Statistics, University of Vermont,
  Burlington,
  VT, 05401
}
\affiliation{
  Vermont Complex Systems Center,
  University of Vermont,
  Burlington,
  VT, 05401
}

\author{
\firstname{Eric M.}
\surname{Clark}
}

\affiliation{
  Computational Story Lab,
  Vermont Advanced Computing Core,
  \& the Department of Mathematics and Statistics, University of Vermont,
  Burlington,
  VT, 05401
}
\affiliation{
  Vermont Complex Systems Center,
  University of Vermont,
  Burlington,
  VT, 05401
}

\author{
\firstname{Suma}
\surname{Desu}
}

\affiliation{
  Center for Computational Engineering,
  Massachusetts Institute of Technology,
  Cambridge,
  MA, 02139
}

\author{
\firstname{Morgan R.}
\surname{Frank}
}

\affiliation{
  Computational Story Lab,
  Vermont Advanced Computing Core,
  \& the Department of Mathematics and Statistics, University of Vermont,
  Burlington,
  VT, 05401
}
\affiliation{
  Vermont Complex Systems Center,
  University of Vermont,
  Burlington,
  VT, 05401
}

\author{
\firstname{Andrew J.}
\surname{Reagan}
}

\affiliation{
  Computational Story Lab,
  Vermont Advanced Computing Core,
  \& the Department of Mathematics and Statistics, University of Vermont,
  Burlington,
  VT, 05401
}
\affiliation{
  Vermont Complex Systems Center,
  University of Vermont,
  Burlington,
  VT, 05401
}

\author{
\firstname{Jake Ryland}
\surname{Williams}
}

\affiliation{
  Computational Story Lab,
  Vermont Advanced Computing Core,
  \& the Department of Mathematics and Statistics, University of Vermont,
  Burlington,
  VT, 05401
}
\affiliation{
  Vermont Complex Systems Center,
  University of Vermont,
  Burlington,
  VT, 05401
}

\author{
\firstname{Lewis}
\surname{Mitchell}
}

\affiliation{
  Computational Story Lab,
  Vermont Advanced Computing Core,
  \& the Department of Mathematics and Statistics, University of Vermont,
  Burlington,
  VT, 05401
}
\affiliation{
  Vermont Complex Systems Center,
  University of Vermont,
  Burlington,
  VT, 05401
}

\author{
\firstname{Kameron Decker}
\surname{Harris}
}

\affiliation{
  Applied Mathematics, 
  University of Washington,
  Lewis Hall \#202, 
  Box 353925, 
  Seattle, 
  WA, 98195.
}

\author{
\firstname{Isabel M.}
\surname{Kloumann}
}

\affiliation{
  Center for Applied Mathematics,
  Cornell University,
  Ithaca,
  NY, 14853.
}

\author{
\firstname{James P.}
\surname{Bagrow}
}

\affiliation{
  Computational Story Lab,
  Vermont Advanced Computing Core,
  \& the Department of Mathematics and Statistics, University of Vermont,
  Burlington,
  VT, 05401
}
\affiliation{
  Vermont Complex Systems Center,
  University of Vermont,
  Burlington,
  VT, 05401
}

\author{
\firstname{Karine}
\surname{Megerdoomian}
}

\affiliation{
  The MITRE Corporation,
  7525 Colshire Drive,
  McLean, 
  VA, 22102
}

\author{
\firstname{Matthew T.}
\surname{McMahon}
}

\affiliation{
  The MITRE Corporation,
  7525 Colshire Drive,
  McLean, 
  VA, 22102
}

\author{
\firstname{Brian F.}
\surname{Tivnan}
}
\email{btivnan@mitre.org}

\affiliation{
  The MITRE Corporation,
  7525 Colshire Drive,
  McLean, 
  VA, 22102
}

\affiliation{
  Vermont Complex Systems Center,
  University of Vermont,
  Burlington,
  VT, 05401
}

\author{
\firstname{Christopher M.}
\surname{Danforth}
}
\email{chris.danforth@uvm.edu}

\affiliation{
  Computational Story Lab,
  Vermont Advanced Computing Core,
  \& the Department of Mathematics and Statistics, University of Vermont,
  Burlington,
  VT, 05401
}
\affiliation{
  Vermont Complex Systems Center,
  University of Vermont,
  Burlington,
  VT, 05401
}

\date{\today}

\begin{abstract}
   
Using human evaluation of 100,000 words spread across 24 corpora in 10
languages diverse in origin and culture, we present evidence of a deep
imprint of human sociality in language, observing that (1) the words
of natural human language possess a universal positivity bias; (2) the
estimated emotional content of words is consistent between languages
under translation; and (3) this positivity bias is strongly
independent of frequency of word usage.
Alongside these general regularities, we describe inter-language
variations in the emotional spectrum of languages which allow us to
rank corpora.
We also show how our word evaluations can be used to construct
physical-like instruments for both real-time and offline measurement
of the emotional content of large-scale texts.

\end{abstract}

\maketitle

Human language---our great social technology---reflects that
which it describes through the stories it allows to be told,
and us, the tellers of those stories.
While language's shaping effect on thinking 
has long been
controversial~\cite{whorf1956a,chomsky1957a,pinker1994a},
we know that
a rich array of metaphor encodes our conceptualizations~\cite{lakoff1980a}, 
word choice reflects our internal motives and immediate social roles~\cite{campbell2003a,newman2003a,pennebaker2011a},
and
the way a language represents the present and future
may condition economic choices~\cite{chen2013a}.

In 1969, Boucher and Osgood framed the Pollyanna Hypothesis: 
a hypothetical, universal positivity bias in human communication~\cite{boucher1969a}.
From a selection of small-scale, cross-cultural studies, 
they marshaled evidence that positive words
are likely more prevalent, more meaningful,
more diversely used, and more readily learned.
However, in being far from an exhaustive, data-driven analysis of 
language---the approach we take here---their 
findings could only be regarded as suggestive.
Indeed, studies of the positivity of isolated words and word stems have produced conflicting
results, some pointing toward a positivity bias~\cite{bradley1999a},
others the opposite~\cite{stone1966a,pennebaker2007a}, though
attempts to adjust for usage frequency tend to recover a positivity signal~\cite{jurafsky2014a}.

To deeply explore the positivity of human language,
we constructed 24 corpora spread across 10 languages
(see Supplementary Online Material).
Our global coverage of linguistically and culturally diverse
languages
includes 
English, 
Spanish,
French,
German,
Brazilian Portuguese,
Korean, 
Chinese (Simplified), 
Russian, 
Indonesian, 
and Arabic.
The sources of our corpora are similarly broad,
spanning 
books~\cite{googlebooks-ngrams2014a},
news outlets,
social media,
the web~\cite{googleweb2006a},
television and movie subtitles,
and music lyrics~\cite{dodds2009a}.
Our work here greatly expands upon our earlier study of
English alone, where we found strong evidence for a usage-invariant
positivity bias~\cite{kloumann2012b}.

We address the social nature of language in two important ways:
(1) we focus on the words people most commonly use,
and
(2) we measure how those same words are received by individuals.
We take word usage frequency as the primary organizing measure
of a word's importance.  
Such a data-driven approach is crucial for both understanding
the structure of language and for creating 
linguistic instruments for principled measurements~\cite{dodds2011e,mitchell2013a}.
By contrast, earlier studies focusing on meaning and emotion
have used `expert' generated word lists, 
and these
fail to statistically match frequency
distributions  of natural
language~\cite{osgood1957a,stone1966a,bradley1999a,pennebaker2007a},
confounding attempts to make claims
about language in general.
For each of our corpora
we selected between 5,000 to 10,000 of the most
frequently used words, 
choosing the exact numbers so that we obtained
approximately 10,000 words for each language.

We then paid native speakers to rate how they
felt in response to individual words on a 9 point scale, with 1 
corresponding to most negative or saddest, 5 to neutral, 
and 9 to most positive or happiest~\cite{bradley1999a,dodds2011e}
(see also Supplementary Online Material).
This happy-sad semantic differential~\cite{osgood1957a}
functions as
a coupling of two standard 5-point Likert scales.
Participants were restricted to certain regions or countries
(for example, Portuguese was rated by residents of Brazil).
Overall, we collected 50 ratings per word for a total
of around 5,000,000 individual human assessments,
and we provide all data sets as part
of the Supplementary Online Material.

\begin{figure}[tbp!]
  \centering
  \includegraphics[width=0.498\textwidth]{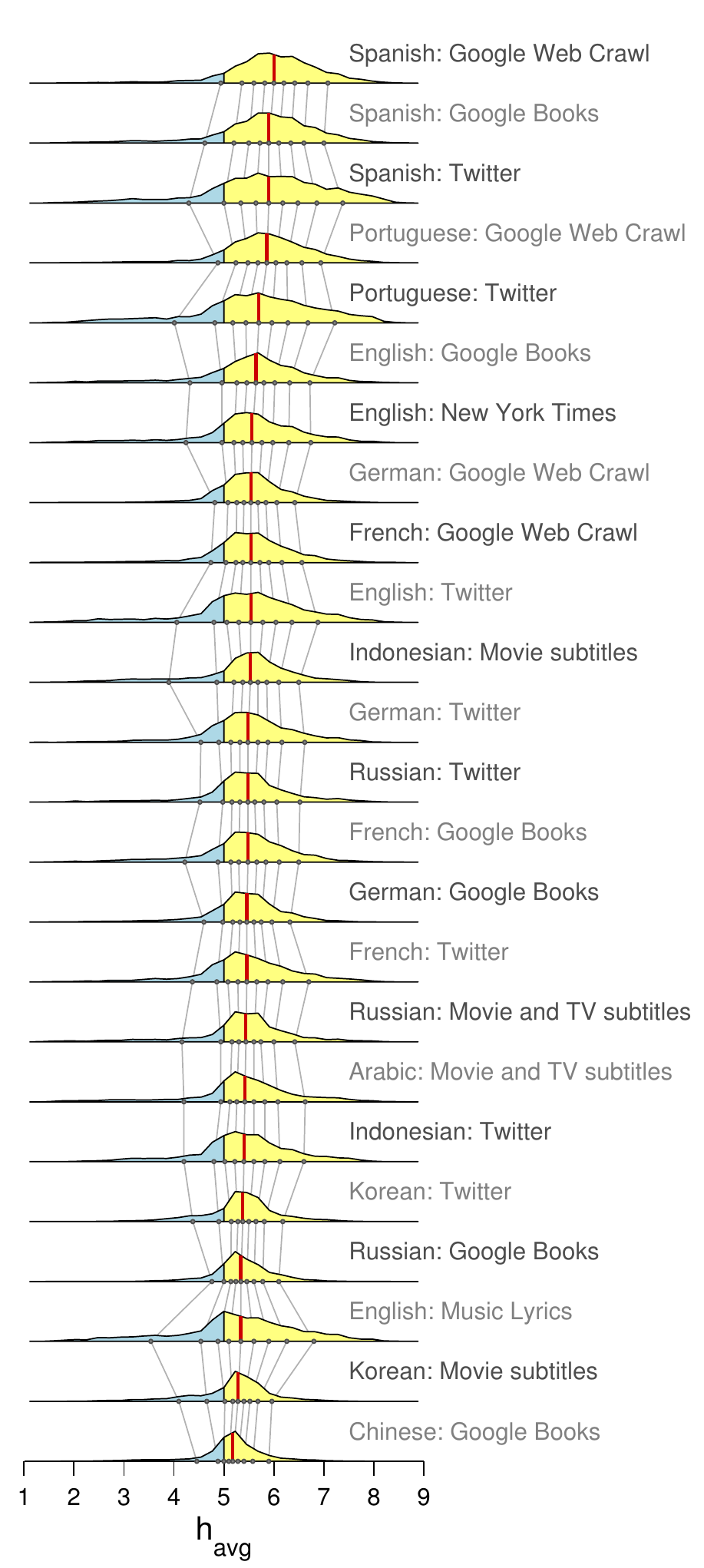}
  \caption{
    Distributions of perceived average word happiness $h_{\rm avg}$ for 
    24 corpora in 10 languages.
    The histograms represent the 5000 most commonly used
    words in each corpora (see Supplementary Online Material for details),
    and native speakers scored words on a 1 to 9 double-Likert scale
    with 
    1 being extremely negative, 
    5 neutral, 
    and 9 extremely positive.
    Yellow indicates positivity ($h_{\rm avg}> 5$) 
    and blue negativity ($h_{\rm avg}< 5$),
    and distributions are
    ordered by increasing median (red vertical line).
    The background grey lines connect deciles of adjacent
    distributions.
    Fig.~\ref{fig:happinessdist_comparison_variance} shows
    the same distributions arranged according to increasing variance.
  }
  \label{fig:mlhap.happinessdist_comparison}
\end{figure}

\begin{figure*}
  \centering
  \includegraphics[width=\textwidth]{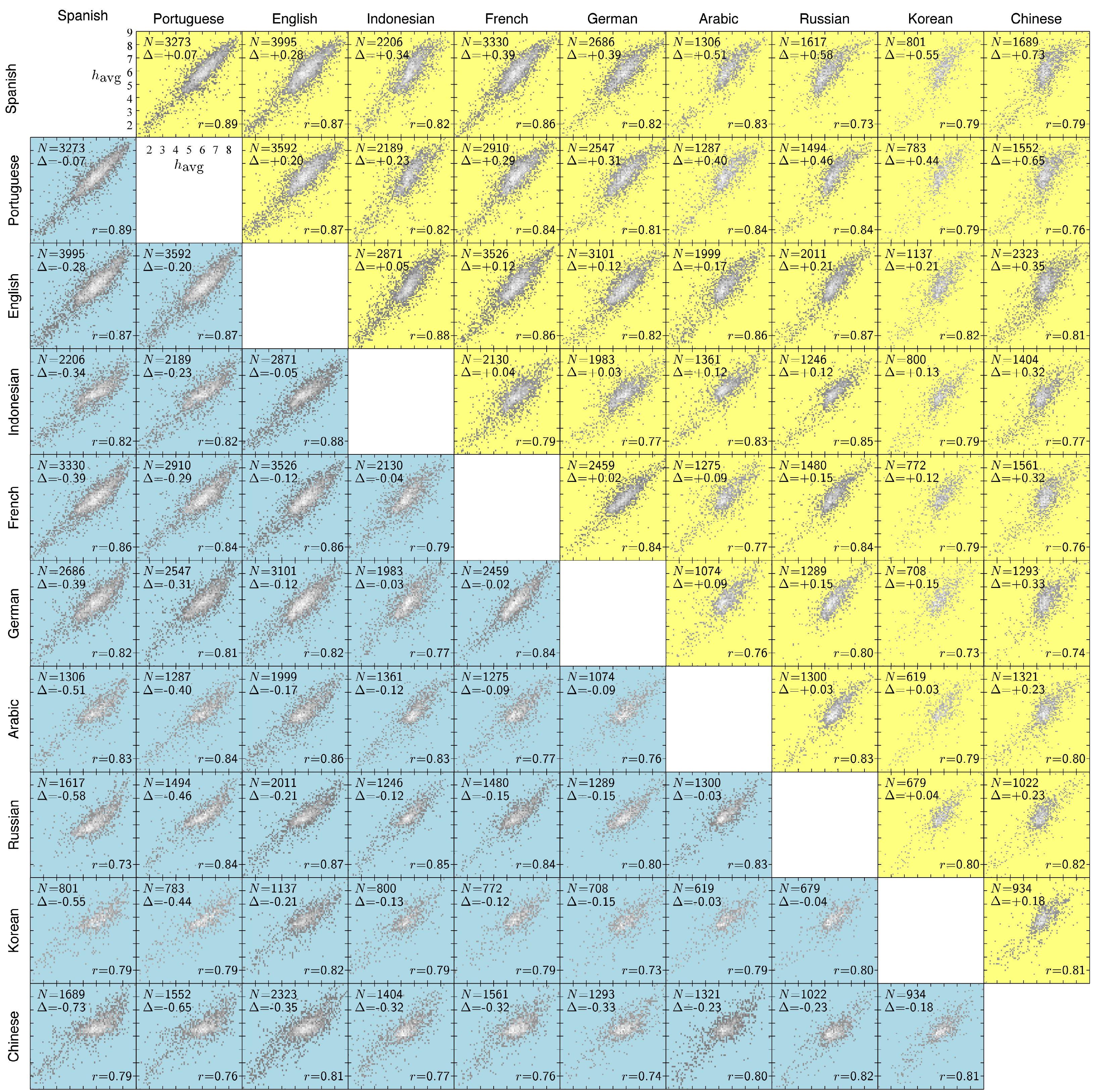}
  \caption{
    Scatter plots of average happiness for words measured in different languages. 
    We order languages from relatively most positive (Spanish)
    to relatively least positive (Chinese); a yellow background indicates 
    the row language is more positive than the column language,
    and a blue background the converse.
    The overall plot matrix is symmetric about the leading diagonal,
    the redundancy allowing for easier comparison between languages.
    In each scatter plot, the key gives 
    the number of translation-stable words for each language
    pair, $N$;
    the average difference in translation-stable word happiness
    between the row language and column language, $\Delta$;
    and
    the Pearson correlation coefficient
    for the regression, $r$.
    All $p$-values are 
    less than $10    and less than $10    Fig.~\ref{fig:mlhap.TranslationValenceCheckboard_hist}
    shows histograms of differences in average happiness
    for translation-stable words.
  }
  \label{fig:mlhap.TranslationValenceCheckboard}
\end{figure*}

\begin{figure*}
  \centering
  \includegraphics[width=\textwidth]{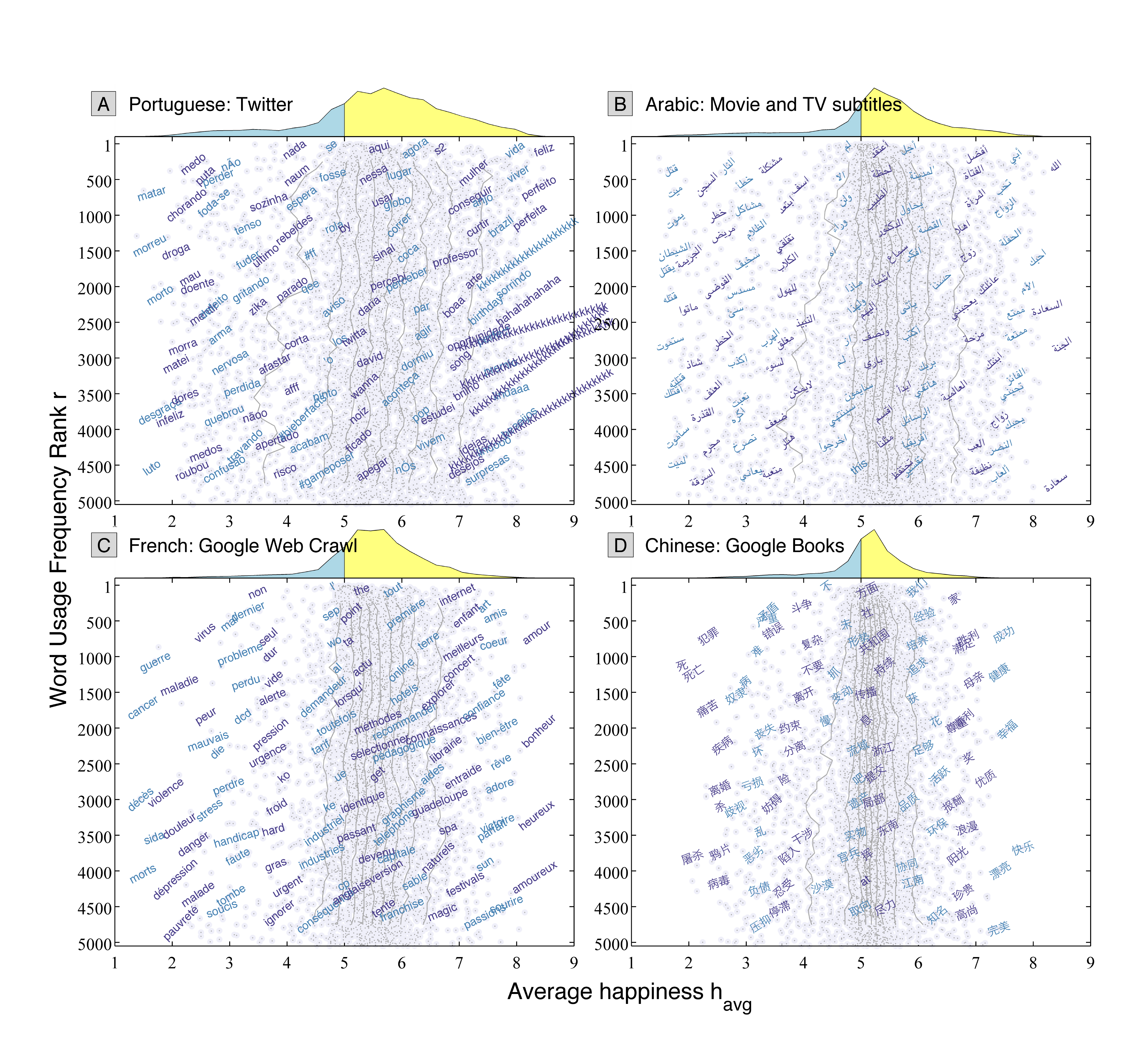}
  \caption{
    Examples of how word happiness varies little
    with usage frequency.
    Above each plot is a histogram of average happiness $h_{\rm avg}$
    for the 5000 most frequently used words in the given corpus, matching
    Fig.~\ref{fig:mlhap.happinessdist_comparison}.
    Each point locates a word by its rank $r$ and average happiness
    $h_{\rm avg}$, and we show some regularly spaced example words.
    The descending gray curves of these jellyfish plots
    indicate deciles for windows of 500 words of
    contiguous usage rank, showing that the overall histogram's form is
    roughly maintained at all scales.
    The `kkkkkk...' words represent laughter in Brazilian Portuguese,
    in the manner of `hahaha...'.
    See
    Fig.~\ref{fig:mlhap.jellyfish_translated} for an English translation,
    Figs.~\ref{fig:mlhap.happinessdist_jellyfish_words_havg_multilanguage001_table1}--\ref{fig:mlhap.happinessdist_jellyfish_words_havg_multilanguage001_table4}
    for all corpora,
    and
    Figs.~\ref{fig:mlhap.happinessdist_jellyfish_words_hstd_multilanguage001_table1}--\ref{fig:mlhap.happinessdist_jellyfish_words_hstd_multilanguage001_table4}
    for the equivalent plots for standard deviation of word happiness
    scores.\\
  }
  \label{fig:mlhap.jellyfish}
\end{figure*}

In Fig.~\ref{fig:mlhap.happinessdist_comparison},
we show distributions of the average happiness scores for
all 24 corpora,
leading to our most general observation
of a clear positivity bias in natural language.
We indicate the above neutral part of each distribution with yellow,
below neutral with blue,
and order the distributions moving upwards
by increasing median (vertical red line).
For all corpora, the median clearly exceeds the neutral score of 5.
The background gray lines connect deciles for each distribution. 
In Fig.~\ref{fig:happinessdist_comparison_variance}, we provide
the same distributions ordered instead by increasing variance.

As is evident from the ordering in
Figs.~\ref{fig:mlhap.happinessdist_comparison}
and~\ref{fig:happinessdist_comparison_variance}, while a positivity
bias is the universal rule, there are minor differences between the
happiness distributions of languages.
For example, Latin American-evaluated corpora (Mexican Spanish and Brazilian
Portuguese) exhibit relatively high medians and, to a lesser degree,
higher variances.
For other languages, we see those with multiple corpora have more
variable medians, and specific corpora are not ordered by median in
the same way across languages (e.g., Google Books has a lower median
than Twitter for Russian, but the reverse is true for German and
English).
In terms of emotional variance, all four English corpora are among the
highest, while Chinese and Russian Google Books seem especially
constrained.

We now examine how individual words themselves
vary in their average happiness score between languages.
Owing to the scale of our corpora, we were compelled to use an online
service, choosing Google Translate.
For each of the 45 language pairs, we translated isolated words from
one language to the other and then back.  
We then found all word pairs that 
(1) were translationally-stable,
meaning the forward and back translation returns the original word,
and (2)
appeared in our corpora for each language.

We provide the resulting comparison between languages at the level of
individual words in Fig.~\ref{fig:mlhap.TranslationValenceCheckboard}.
We use the mean of each language's word happiness distribution derived
from their merged corpora
to generate a rough overall ordering, acknowledging that frequency of
usage is no longer meaningful, and moreover is not relevant as we are
now investigating the properties of individual words.
Each cell shows a heat map comparison with word density increasing as
shading moves from gray to white.
The background colors reflect the ordering of each pair of languages,
yellow if the row language had a higher average happiness than
the column language, and blue for the reverse.
In each cell, we display the number of translation-stable words between language
pairs, $N$, along with the difference in average word happiness, $\Delta$,
where each word is equally weighted.

A linear relationship is clear for each language-language
comparison, and is supported by Pearson's correlation
coefficient $r$ 
being in the range 0.73 to 0.89 ($p$-value $< 10^{-118}$ across all
pairs; see Fig.~\ref{fig:mlhap.TranslationValenceCheckboard}
and Tabs.~\ref{tab:mhl.translationlinearfits},
\ref{tab:mhl.translation_pearson},
and
\ref{tab:mhl.translation_spearman}).
Overall, this strong agreement between languages, previously observed
on a small scale for a Spanish-English translation~\cite{redondo2007a},
suggests that approximate estimates of word happiness for unscored
languages could be generated with no expense from our existing
data set.
Some words will of course translate unsatisfactorily, with the
dominant
meaning changing between languages.
For example `lying' in English, most readily interpreted as speaking
falsehoods by our participants, translates to `acostado' in Spanish,
meaning recumbent.
Nevertheless, happiness scores obtained by translation will be serviceable for 
purposes where the effects of many different words are incorporated.
(See the Supplementary Online Material for links to an interactive visualization of
Fig.~\ref{fig:mlhap.TranslationValenceCheckboard}.)

Stepping back from examining inter-language robustness,
we return to a more detailed exploration of the 
rich structure of each corpus's happiness distribution.
In Fig.~\ref{fig:mlhap.jellyfish}, we show
how average word happiness $\havgfn$ is largely
independent of word usage frequency for four example corpora.
We first plot usage frequency rank $r$ of the 5000 most frequently used
words 
as a function of their average happiness score, $\havgfn$ (background
dots), along with some example evenly-spaced words.
(We note that words at the extremes of the happiness scale
are ones evaluators agreed upon strongly, while words near neutral 
range from being clearly neutral (e.g., $\havgword{the}$=4.98)
to contentious with high standard deviation~\cite{kloumann2012b}.)
We then compute deciles for contiguous sets of 500 words,
sliding this window through rank $r$.  
These deciles form the vertical strands.
We overlay randomly chosen, equally-spaced example words 
to give a sense of each corpus's emotional texture.

We chose the four example corpora 
shown in Fig.~\ref{fig:mlhap.jellyfish} to be disparate in nature,
covering
diverse languages (French, Egyptian Arabic, Brazilian Portuguese, and
Chinese), 
regions of the world (Europe, the Middle East, South America,
and Asia), and texts (Twitter, movies and television,
the Web~\cite{googleweb2006a}, and books~\cite{googlebooks-ngrams2014a}).
In the Supplementary Online Material, we show all 24 corpora
yield similar plots 
(see Figs.~\ref{fig:mlhap.happinessdist_jellyfish_words_havg_multilanguage001_table1}--\ref{fig:mlhap.happinessdist_jellyfish_words_havg_multilanguage001_table4}
and English translated versions, Figs.~\ref{fig:mlhap.happinessdist_jellyfish_words_havg_multilanguage002_table1}--\ref{fig:mlhap.happinessdist_jellyfish_words_havg_multilanguage002_table4}).
We also show how the 
standard deviation for word happiness exhibits
an approximate self-similarity
(Figs.~\ref{fig:mlhap.happinessdist_jellyfish_words_hstd_multilanguage001_table1}--\ref{fig:mlhap.happinessdist_jellyfish_words_hstd_multilanguage001_table4}
and their translations, Figs.~\ref{fig:mlhap.happinessdist_jellyfish_words_hstd_multilanguage002_table1}--\ref{fig:mlhap.happinessdist_jellyfish_words_hstd_multilanguage002_table4}).

Across all corpora, we observe visually that the deciles tend to stay fixed
or move slightly toward the negative, with
some expected fragility at the 10\% and 90\% levels (due to the
distributions' tails), indicating
that each corpus's overall happiness distribution approximately 
holds independent of word usage.
In Fig.~\ref{fig:mlhap.jellyfish}, for example, we see that 
both the Brazilian Portuguese and French examples show a small
shift to the negative for increasingly rare words, while there
is no visually clear trend for the Arabic and Chinese cases.
Fitting $h_{\textnormal{avg}} = \alpha r + \beta$ typically
returns $\alpha$ on the order of -1$\times$$10^{-5}$ suggesting 
$\havgfn$ decreases 0.1 per 10,000 words.
For standard deviations of happiness scores
(Figs.~\ref{fig:mlhap.happinessdist_jellyfish_words_hstd_multilanguage001_table1}--\ref{fig:mlhap.happinessdist_jellyfish_words_hstd_multilanguage001_table4}),
we find a similarly weak drift toward higher values
for increasingly rare words
(see Tabs.~\ref{tab:mhl.havgcorr} and \ref{tab:mhl.hstdcorr} for
correlations and linear fits for $\havgfn$ and $\hstdfn$ as a function
of word rank $r$ for all corpora).
We thus find that, to first order, not just the positivity bias, but the happiness 
distribution itself applies for common words and rare words alike,
revealing an unexpected addition to the many well known
scalings found in natural language, famously exemplified by Zipf's law~\cite{zipf1949a}.

\begin{figure*}
  \centering
  \includegraphics[width=\textwidth]{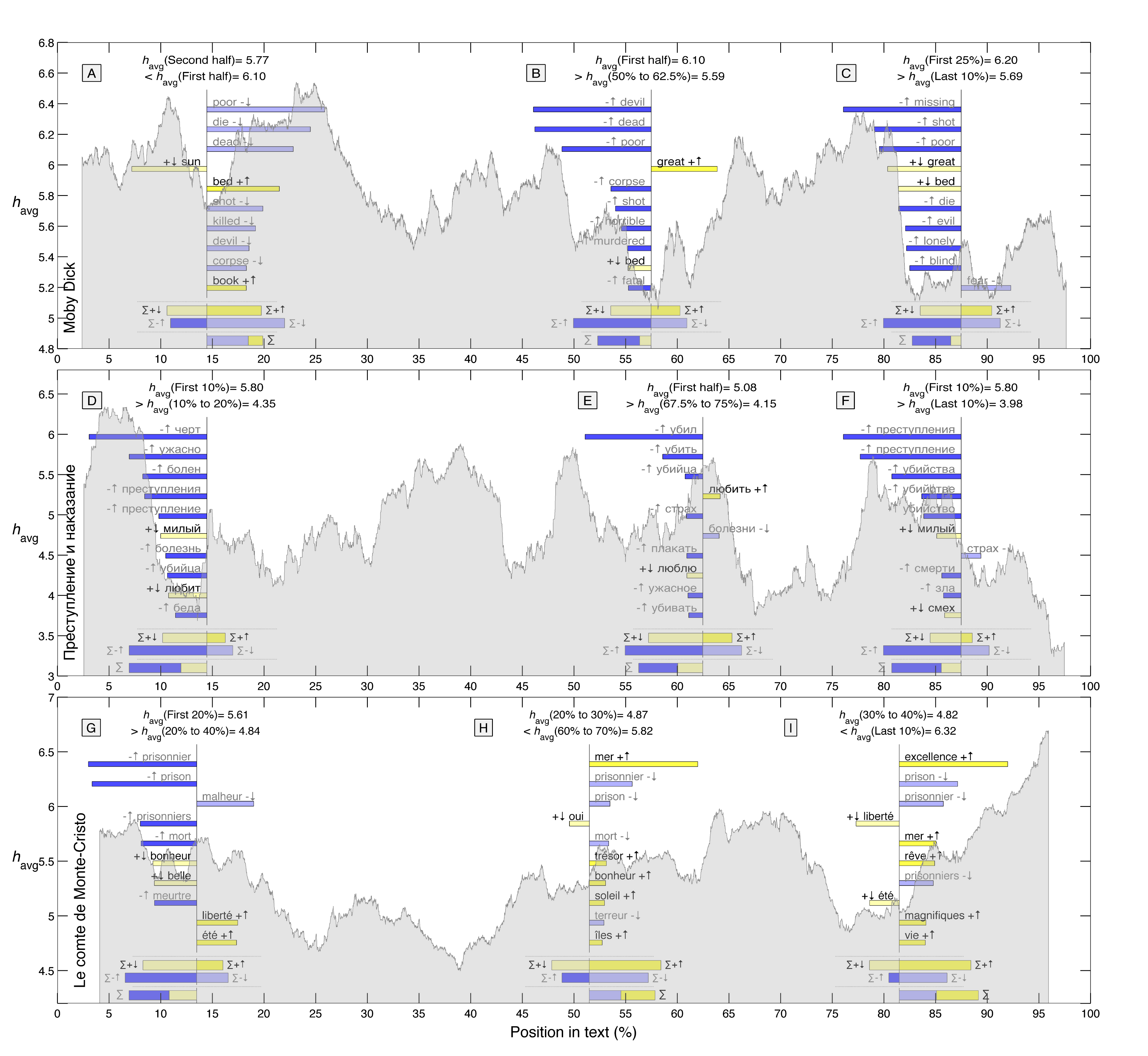}
  \caption{
            Emotional time series for three great 19th century works of
    literature:
    Melville's Moby Dick,
    Dostoyevsky's Crime and Punishment,
    and 
    Dumas' Count of Monte Cristo.
    Each point represents the language-specific happiness score for a window of 10,000
    words (converted to lowercase), with the window translated throughout the work.
    The overlaid word shifts show example comparisons between
    different
    sections of each work.
    Word shifts
    indicate which words contribute the most toward and
    against
    the change in average happiness between two texts
    (see pp.~\pageref{page:mhl.startwordshiftdescription}--\pageref{page:mhl.endwordshiftdescription}).
    While a robust instrument in general, 
    we acknowledge the hedonometer's potential failure for individual words
    both due to language evolution and words possessing
    more than one meaning.
    While a robust instrument in general, 
    we acknowledge the hedonometer's potential failure for individual words
    both due to language evolution and words possessing
    more than one meaning.
    For Moby Dick, we excluded `cried' and `cry' (to speak loudly
    rather than weep)
    and `Coffin' (surname, still common on Nantucket).  
    Such alterations, which can
    be done on a case by case basis, do not
    noticeably change the overall happiness curves while leaving the 
    word shifts more informative.
    We provide links to online, interactive versions of these time series
    in the Supplementary Online Information.
  }
  \label{fig:mlhap.measurementexamples}
\end{figure*}

In constructing language-based instruments for measuring
expressed happiness, such as our hedonometer~\cite{dodds2011e},
this frequency independence allows for a way to 
`increase the gain' in a way resembling that of standard physical instruments.
Moreover, we have earlier demonstrated the robustness
of our hedonometer for the English language,
showing, for example that measurements derived from Twitter correlate strongly
with Gallup well-being polls and related indices at the state and city 
level for the United States~\cite{mitchell2013a}.

Here, we provide an illustrative use of our hedonometer
in the realm of literature,
inspired by Vonnegut's shapes of stories~\cite{vonnegut2005a,vonnegut2010a}.
In Fig.~\ref{fig:mlhap.measurementexamples},
we show `happiness time series' for three
famous works of literature, evaluated in their original languages
English, Russian, and French:
\textbf{A.} Melville's Moby Dick~\cite{gutenberg2013a},
\textbf{B.} Dostoyevsky's Crime and Punishment~\cite{dostoyevsky2013a},
and 
\textbf{C.} Dumas' Count of Monte Cristo~\cite{gutenberg2013a}.
We slide a 10,000-word window through each work,
computing the average happiness
using a `lens' for the hedonometer in the following manner.
We capitalize on our instrument's tunablility
to obtain a strong signal by excluding all words for which $3 <
\havgfn < 7$, i.e., we keep words residing in the tails of each
distribution~\cite{dodds2011e}.
Denoting a given lens by its corresponding set of allowed words $L$,
we estimate the happiness score of any text $T$
as 
$\havg{T} =\sum_{w \in L}f_{w} \havg{w}/\sum_{w \in L}f_{w}$
where $f_{w}$ is the frequency of word $w$ in $T$~\cite{dodds2009b}.

The three resulting happiness time series provide interesting,
detailed views of each work's narrative trajectory
revealing numerous peaks and troughs throughout,
at times clearly dropping below neutral.
Both Moby Dick and Crime and Punishment
end on low notes, whereas the Count of Monte Cristo 
culminates with a rise in positivity, accurately reflecting the
finishing arcs of all three.
The `word shifts' overlaying the time series
compare two distinct regions of each work,
showing how changes in word abundances lead to
overall shifts in average happiness.
Such word shifts are essential tests of any sentiment
measurement, and are made possible by the linear form of our
instrument~\cite{dodds2009b,dodds2011e}
(see
pp.~\pageref{page:mhl.startwordshiftdescription}--\pageref{page:mhl.endwordshiftdescription}
in the Supplementary Online Material for a full explanation).
As one example, the third word shift for Moby Dick shows
why the average happiness of the last 10\% of the book
is well below that of the first 25\%.  
The major contribution is an increase in relatively negative
words including `missing', `shot', `poor', `die', and `evil'.
We include full diagnostic versions of all word shifts
in Figs.~\ref{fig:universal_wordshift500_figuniversal_wordshift_hli_moby_dick001}--\ref{fig:universal_wordshift500_figuniversal_wordshift_hli_count_of_monte_cristo003}.

By adjusting the lens, many other related time series can be formed 
such as those produced by focusing on only positive or negative words.
Emotional variance as a function of text position can also be readily
extracted. In the Supplementary Online Material, 
we provide links to online, interactive versions 
of these graphs where different lenses and 
regions of comparisons may be easily explored.
Beyond this example tool we have created here for the digital
humanities and our hedonometer for measuring population well-being,
the data sets we have generated for the present study
may be useful in creating a great variety of 
language-based instruments for assessing emotional expression.

Overall, our major scientific finding is that 
when experienced in isolation and weighted properly according to
usage, words---the atoms of human language---present
an emotional spectrum with a universal, self-similar positive bias.
We emphasize that this apparent linguistic encoding of our social nature is a system level
property, and in no way asserts all natural texts will skew positive (as exemplified
by certain passages of the three works in
Fig.~\ref{fig:mlhap.measurementexamples}),
or diminishes the salience of negative states~\cite{forgas2013a}.
Nevertheless, a general positive bias points towards
a positive social evolution, and may be linked
to the gradual if haphazard trajectory of modern civilization
toward greater human rights and decreases in violence~\cite{pinker2011a}.
Going forward, our word happiness assessments should be 
periodically repeated, and
carried out for new languages, tested on different
demographics, and expanded to phrases,
both for the improvement of hedonometric instruments 
and to chart the dynamics of our
collective social self.

\revtexonly{
  The authors acknowledge 
  I.~Ramiscal, C.~Burke, P.~Carrigan, M.~Koehler, 
  and Z.~Henscheid, in part for their roles
  in developing \href{http://www.hedonometer.org}{hedonometer.org}.
  The authors are also grateful
  for conversations with
  F.~Henegan, 
  A.~Powers,
  and N.~Atkins.
  PSD was
  supported by NSF CAREER Award \# 0846668.  
}

\clearpage

\newwrite\tempfile
\immediate\openout\tempfile=startsupp.txt
\immediate\write\tempfile{\thepage}
\immediate\closeout\tempfile

\setcounter{page}{1}
\renewcommand{\thepage}{S\arabic{page}}
\renewcommand{\thefigure}{S\arabic{figure}}
\renewcommand{\thetable}{S\arabic{table}}
\setcounter{figure}{0}
\setcounter{table}{0}
\section*{Supplementary Online Material}

\subsection*{Online, interactive visualizations:}

Spatiotemporal hedonometric measurements of Twitter across
all 10 languages can be explored at 
\href{http://www.hedonometer.org}{hedonometer.org}.

We provide the following resources online at 
\url{http://www.uvm.edu/~storylab/share/papers/dodds2014a/}.

\begin{itemize}
\item 
  Example scripts for parsing and measuring average happiness
  scores for texts;
\item 
  D3 and Matlab scripts for generating word shifts;
\item 
  Visualizations for exploring translation-stable word pairs
  across languages;
\item
  Interactive time series for Moby Dick, Crime and Punishment,
  the Count of Monte Cristo, and other works of literature.
\end{itemize}

\begin{table*}[tbp!]
  \centering
  \rowcolors{1}{verylightgrey}{white}
  \begin{tabular}{|l|c|c|}
    \hline
    \hline
    Corpus: & \# Words &  Reference(s) \\
    \hline
    English: Twitter  & 5000 & \cite{twitterapi,dodds2011e} \\
    English: Google Books Project & 5000 & \cite{googlebooks-ngrams2014a,michel2011a}\\
    English: The New York Times & 5000 & \cite{nytimescorpus2008a} \\
    English: Music lyrics & 5000 & \cite{dodds2009b} \\
    Portuguese: Google Web Crawl & 7133 & \cite{googleweb2006a}\\
    Portuguese: Twitter  & 7119 & \cite{twitterapi} \\
    Spanish: Google Web Crawl & 7189 & \cite{googleweb2006a}\\
    Spanish: Twitter  & 6415 & \cite{twitterapi} \\
    Spanish: Google Books Project & 6379 & \cite{googlebooks-ngrams2014a,michel2011a}\\
    French: Google Web Crawl & 7056 & \cite{googleweb2006a}\\
    French: Twitter  & 6569 & \cite{twitterapi} \\
    French: Google Books Project & 6192 & \cite{googlebooks-ngrams2014a,michel2011a}\\
    Arabic: Movie and TV subtitles & 9999 & The MITRE Corporation \\
    Indonesian: Twitter & 7044 & \cite{twitterapi} \\
    Indonesian: Movie subtitles & 6726 & The MITRE Corporation \\
    Russian: Twitter  & 6575 & \cite{twitterapi} \\
    Russian: Google Books Project & 5980 & \cite{googlebooks-ngrams2014a,michel2011a}\\
    Russian: Movie and TV subtitles & 6186 & \cite{googleweb2006a}\\
    German: Google Web Crawl & 6902 & \cite{googleweb2006a}\\
    German: Twitter  & 6459 & \cite{twitterapi} \\
    German: Google Books Project & 6097 & \cite{googlebooks-ngrams2014a,michel2011a}\\
    Korean: Twitter & 6728 & \cite{twitterapi} \\
    Korean: Movie subtitles & 5389 & The MITRE Corporation \\
    Chinese: Google Books Project & 10000 & \cite{googlebooks-ngrams2014a,michel2011a} \\
    \hline
    \hline
  \end{tabular}
  \caption{
    Sources for all corpora.
  }
  \label{tab:mhl.corpora}
\end{table*}

\begin{table}
  \centering
  \begin{tabular}{rl}
    \hline
    \hline
    English & United States of America, India \\
    German & Germany \\
    Indonesian & Indonesia \\
    Russian & Russia \\
    Arabic & Egypt \\
    French & France \\
    Spanish & Mexico  \\
    Portuguese & Brazil \\
    Simplified Chinese &        China    \\
    Korean &  Korea, United States of America \\
    \hline
    \hline
  \end{tabular}
  \caption{
    Main country of location for participants.
  }
  \label{tab:we}
\end{table}

\subsection*{Corpora}

We used the services of Appen Butler Hill (\url{http://www.appen.com})
for all word evaluations excluding English, for which we had
earlier employed Mechanical Turk (\url{https://www.mturk.com/}~\cite{kloumann2012b}).

English instructions were translated to all other languages
and given to participants along with survey questions,
and an example of the English instruction page is below.
Non-english language experiments were conducted through 
a custom interactive website built by Appen Butler Hill,
and all participants were required to pass a stringent 
oral proficiency test in their own language.  

\begin{myframe}
  \includegraphics[width=0.85\columnwidth]{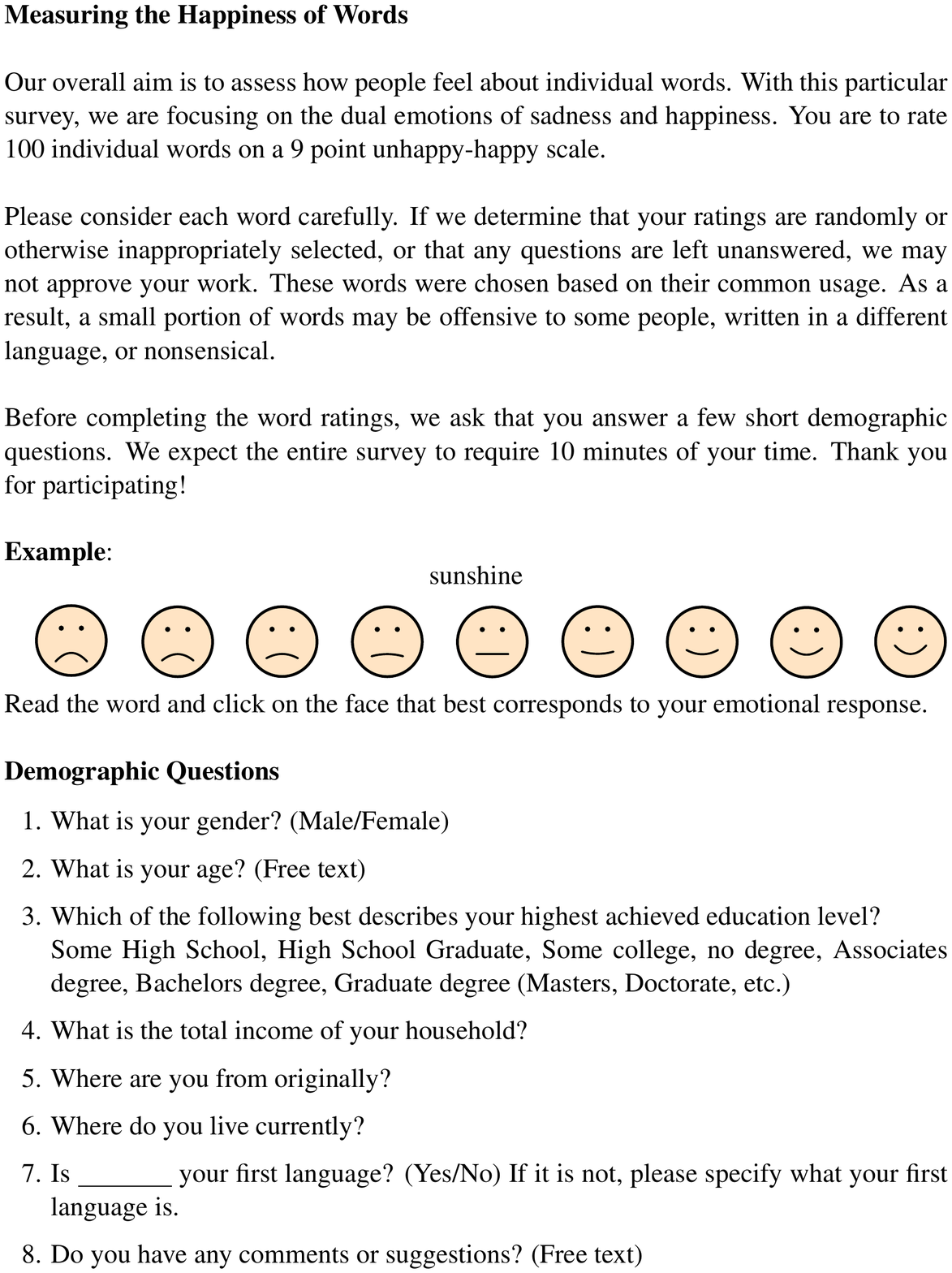}
\end{myframe}

Sizes and sources for our 24 corpora
are given in Tab.~\ref{tab:mhl.corpora}.

We used Mechanical Turk to obtain evaluations of the four English corpora~\cite{kloumann2012b}.
For all non-English assessments,
we contracted the translation services company Appen-Butler Hill.
For each language, participants were required 
to be native speaker, to have grown up in the country where the language is spoken,
and to pass a strenuous online aural comprehension test.

\subsection*{Notes on corpus generation}

There is no single, principled way to merge corpora
to create an ordered list of words for a given language.
For example, it is impossible to weight the most commonly used
words in the New York Times against those of Twitter.
Nevertheless, we are obliged to choose some method for doing so
to facilitate comparisons across languages and for the purposes
of building adaptable linguistic instruments.

For each language, we created a single quasi-ranked word list
by finding the smallest integer $r$ such that
the union of all words with rank $\le r$ in at least
one corpus formed a set of at least 10,000 words.

For Twitter, we first checked if a string contains at least one valid
utf8 letter,
discarding if not.
Next we filtered out strings containing invisible
control characters, as these symbols can be problematic.
We ignored all strings that start with $<$ and end with $>$ (generally
html code).
We ignored strings with a leading @ or \&, or either preceded with
standard punctuation (e.g., Twitter ID's), but kept hashtags.
We also
removed all strings starting with www. or http: or end in .com (all
websites).
We stripped the remaining strings of standard punctuation, and
we replaced all double quotes ('') by single quotes (').
Finally, we converted all Latin alphabet letters to lowercase.

A simple example of this tokenization process would be:
\begin{center}
  \begin{tabular}{lr}
    \hline 
    \hline 
    Term & count \\
    \hline 
    love & 10 \\
    LoVE & 5 \\
    love! & 2 \\
    \#love & 3 \\
    .love & 2 \\
    @love & 1 \\
    love87 & 1 \\
    \hline 
    \hline 
  \end{tabular}
  $\rightarrow$
  \begin{tabular}{lr}
    \hline 
    \hline 
    Term & count \\
    \hline 
    love & 19 \\
    \#love & 3 \\
    love87 & 1 \\
    \hline 
    \hline 
  \end{tabular}
\end{center}
The term `@love' is discarded, and all other terms
map to either `love' or `love87'.

\begin{figure}[tbp!]
  \centering
  \includegraphics[width=.5\textwidth]{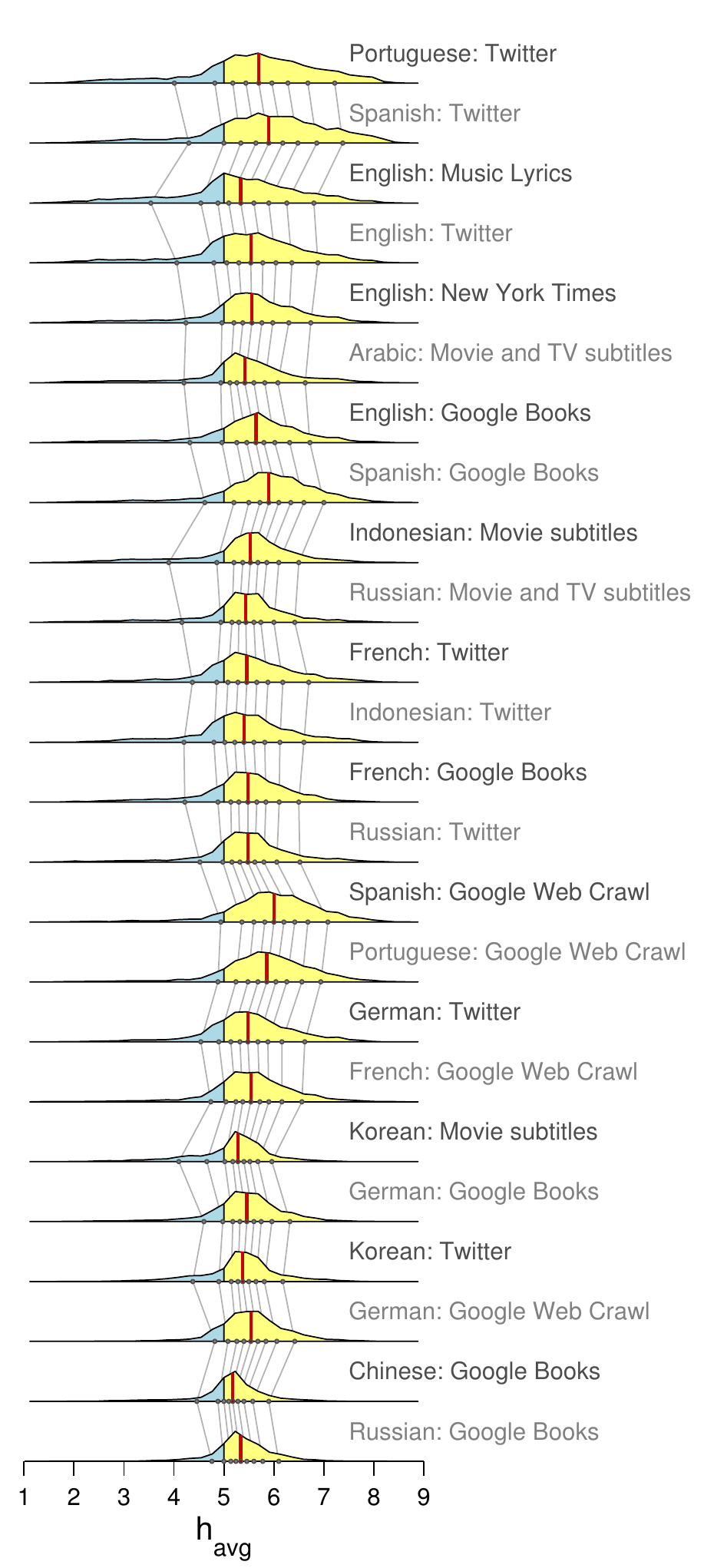}
  \caption{
    The same average happiness distributions shown in
    Fig.~\ref{fig:mlhap.happinessdist_comparison}
    re-ordered by increasing variance.
    Yellow indicates above neutral ($\havgfn = 5$),
    blue below neutral,
    red vertical lines mark each distribution's median, 
    and the gray background lines connect the deciles
    of adjacent distributions.
  }
  \label{fig:happinessdist_comparison_variance}
\end{figure}

\begin{table*}
  \centering
  \rowcolors{1}{verylightgrey}{white}
  \begin{tabular}{|r|c|c|c|c|c|c|c|c|c|c|}
    \hline
    \hline
    & Spanish & Portuguese & English & Indonesian & French & German & Arabic & Russian & Korean & Chinese \\
    \hline
    Spanish & 1.00, 0.00 & 1.01, 0.03 & 1.06, -0.07 & 1.22, -0.88 & 1.11, -0.24 & 1.22, -0.84 & 1.13, -0.22 & 1.31, -1.16 & 1.60, -2.73 & 1.58, -2.30 \\
    Portuguese & 0.99, -0.03 & 1.00, 0.00 & 1.04, -0.03 & 1.22, -0.97 & 1.11, -0.33 & 1.21, -0.86 & 1.09, -0.08 & 1.26, -0.95 & 1.62, -2.92 & 1.58, -2.39 \\
    English & 0.94, 0.06 & 0.96, 0.03 & 1.00, 0.00 & 1.13, -0.66 & 1.06, -0.23 & 1.16, -0.75 & 1.05, -0.10 & 1.21, -0.91 & 1.51, -2.53 & 1.47, -2.10 \\
    Indonesian & 0.82, 0.72 & 0.82, 0.80 & 0.88, 0.58 & 1.00, 0.00 & 0.92, 0.48 & 0.99, 0.06 & 0.89, 0.71 & 1.02, 0.04 & 1.31, -1.53 & 1.33, -1.42 \\
    French & 0.90, 0.22 & 0.90, 0.30 & 0.94, 0.22 & 1.09, -0.52 & 1.00, 0.00 & 1.08, -0.44 & 0.99, 0.12 & 1.12, -0.50 & 1.37, -1.88 & 1.40, -1.77 \\
    German & 0.82, 0.69 & 0.83, 0.71 & 0.86, 0.65 & 1.01, -0.06 & 0.92, 0.41 & 1.00, 0.00 & 0.91, 0.61 & 1.07, -0.25 & 1.29, -1.44 & 1.32, -1.36 \\
    Arabic & 0.88, 0.19 & 0.92, 0.08 & 0.95, 0.10 & 1.12, -0.80 & 1.01, -0.12 & 1.10, -0.68 & 1.00, 0.00 & 1.12, -0.63 & 1.40, -2.14 & 1.43, -2.01 \\
    Russian & 0.76, 0.88 & 0.80, 0.75 & 0.83, 0.75 & 0.98, -0.04 & 0.89, 0.45 & 0.93, 0.24 & 0.89, 0.56 & 1.00, 0.00 & 1.26, -1.39 & 1.25, -1.05 \\
    Korean & 0.62, 1.70 & 0.62, 1.81 & 0.66, 1.67 & 0.77, 1.17 & 0.73, 1.37 & 0.78, 1.12 & 0.71, 1.53 & 0.79, 1.10 & 1.00, 0.00 & 0.98, 0.28 \\
    Chinese & 0.63, 1.46 & 0.63, 1.51 & 0.68, 1.43 & 0.75, 1.07 & 0.71, 1.26 & 0.76, 1.03 & 0.70, 1.41 & 0.80, 0.84 & 1.02, -0.29 & 1.00, 0.00 \\
    \hline
    \hline
  \end{tabular}
  \caption{
    Reduced Major Axis (RMA) regression fits for row language as a linear
    function of the column language:
    $
    \havgfn^{\textnormal{(row)}}(w)
    =
    m \havgfn^{\textnormal{(column)}}(w)
    +
    c
    $
    where $w$ indicates a translation-stable word.
    Each entry in the table contains the coefficient pair $m$ and $c$.
    See the scatter plot tableau of Fig.~\ref{fig:mlhap.TranslationValenceCheckboard}
    for further details on all language-language comparisons.
    We use RMA regression, also known as
    Standardized Major Axis linear regression,
    because of its accommodation of errors in both variables.
      }
  \label{tab:mhl.translationlinearfits}
\end{table*}

\begin{table*}
  \centering
  \rowcolors{1}{verylightgrey}{white}
  \begin{tabular}{|r|c|c|c|c|c|c|c|c|c|c|}
    \hline
    \hline
    & Spanish & Portuguese & English & Indonesian & French & German & Arabic & Russian & Korean & Chinese \\
    \hline
    Spanish & 1.00 & 0.89 & 0.87 & 0.82 & 0.86 & 0.82 & 0.83 & 0.73 & 0.79 & 0.79 \\
    Portuguese & 0.89 & 1.00 & 0.87 & 0.82 & 0.84 & 0.81 & 0.84 & 0.84 & 0.79 & 0.76 \\
    English & 0.87 & 0.87 & 1.00 & 0.88 & 0.86 & 0.82 & 0.86 & 0.87 & 0.82 & 0.81 \\
    Indonesian & 0.82 & 0.82 & 0.88 & 1.00 & 0.79 & 0.77 & 0.83 & 0.85 & 0.79 & 0.77 \\
    French & 0.86 & 0.84 & 0.86 & 0.79 & 1.00 & 0.84 & 0.77 & 0.84 & 0.79 & 0.76 \\
    German & 0.82 & 0.81 & 0.82 & 0.77 & 0.84 & 1.00 & 0.76 & 0.80 & 0.73 & 0.74 \\
    Arabic & 0.83 & 0.84 & 0.86 & 0.83 & 0.77 & 0.76 & 1.00 & 0.83 & 0.79 & 0.80 \\
    Russian & 0.73 & 0.84 & 0.87 & 0.85 & 0.84 & 0.80 & 0.83 & 1.00 & 0.80 & 0.82 \\
    Korean & 0.79 & 0.79 & 0.82 & 0.79 & 0.79 & 0.73 & 0.79 & 0.80 & 1.00 & 0.81 \\
    Chinese & 0.79 & 0.76 & 0.81 & 0.77 & 0.76 & 0.74 & 0.80 & 0.82 & 0.81 & 1.00 \\
    \hline
    \hline
  \end{tabular}
  \caption{
    Pearson correlation coefficients for translation-stable words for
    all language pairs.
    All $p$-values are $<10^{-118}$.
    These values are included in
    Fig.~\ref{fig:mlhap.TranslationValenceCheckboard}
    and reproduced here for to facilitate comparison.
  }
  \label{tab:mhl.translation_pearson}
\end{table*}

\begin{table*}
  \centering
  \rowcolors{1}{verylightgrey}{white}
  \begin{tabular}{|r|c|c|c|c|c|c|c|c|c|c|}
    \hline
    \hline
    & Spanish & Portuguese & English & Indonesian & French & German & Arabic & Russian & Korean & Chinese \\
    \hline
    Spanish & 1.00 & 0.85 & 0.83 & 0.77 & 0.81 & 0.77 & 0.75 & 0.74 & 0.74 & 0.68 \\
    Portuguese & 0.85 & 1.00 & 0.83 & 0.77 & 0.78 & 0.77 & 0.77 & 0.81 & 0.75 & 0.66 \\
    English & 0.83 & 0.83 & 1.00 & 0.82 & 0.80 & 0.78 & 0.78 & 0.81 & 0.75 & 0.70 \\
    Indonesian & 0.77 & 0.77 & 0.82 & 1.00 & 0.72 & 0.72 & 0.76 & 0.77 & 0.71 & 0.71 \\
    French & 0.81 & 0.78 & 0.80 & 0.72 & 1.00 & 0.80 & 0.67 & 0.79 & 0.71 & 0.64 \\
    German & 0.77 & 0.77 & 0.78 & 0.72 & 0.80 & 1.00 & 0.69 & 0.76 & 0.64 & 0.62 \\
    Arabic & 0.75 & 0.77 & 0.78 & 0.76 & 0.67 & 0.69 & 1.00 & 0.74 & 0.69 & 0.68 \\
    Russian & 0.74 & 0.81 & 0.81 & 0.77 & 0.79 & 0.76 & 0.74 & 1.00 & 0.70 & 0.66 \\
    Korean & 0.74 & 0.75 & 0.75 & 0.71 & 0.71 & 0.64 & 0.69 & 0.70 & 1.00 & 0.71 \\
    Chinese & 0.68 & 0.66 & 0.70 & 0.71 & 0.64 & 0.62 & 0.68 & 0.66 & 0.71 & 1.00 \\
    \hline
    \hline
  \end{tabular}
  \caption{
    Spearman correlation coefficients for translation-stable words.
    All $p$-values are $<10^{-82}$.
  }
  \label{tab:mhl.translation_spearman}
\end{table*}

\begin{figure*}
  \centering
  \includegraphics[width=\textwidth]{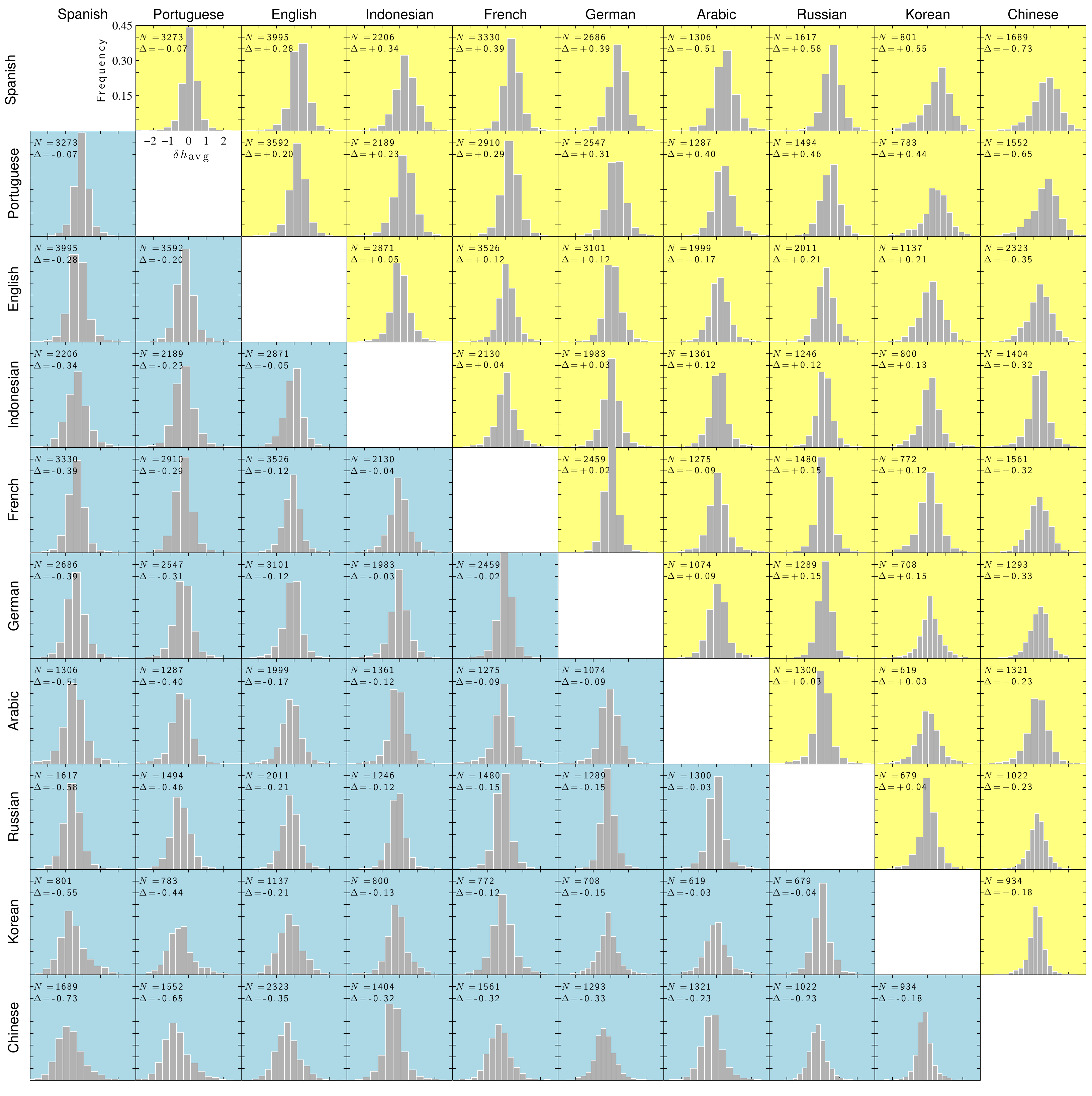}
  \caption{
    Histograms of the change in average happiness for
    translation-stable words between each language pair,
    companion to Fig.~\ref{fig:mlhap.TranslationValenceCheckboard}
    given in the main text.
    The largest deviations correspond to strong changes 
    in a word's perceived primary meaning (e.g., `lying' and `acostado').
    As per Fig.~\ref{fig:mlhap.TranslationValenceCheckboard},
    the inset quantities are $N$, the number of translation-stable words,
    and $\Delta$ is the average difference in translation-stable word happiness
    between the row language and column language.
  }
  \label{fig:mlhap.TranslationValenceCheckboard_hist}
\end{figure*}

\begin{table*}[htbp!]
  \centering
  \rowcolors{1}{verylightgrey}{white}
  \begin{tabular}{|l|c|c|c|c|c|c|}
    \hline
    \hline
    Language: Corpus & $\rho_{\rm p}$ & $p$-value& $\rho_{\rm s}$ & $p$-value & $\alpha$ & $\beta$ \\
    \hline
    Spanish: Google Web Crawl & -0.114 & 3.38$\times$$10^{-22}$ & -0.090 & 1.85$\times$$10^{-14}$ & -5.55$\times$$10^{-5}$ & 6.10 \\
    Spanish: Google Books & -0.040 & 1.51$\times$$10^{-3}$ & -0.016 & 1.90$\times$$10^{-1}$ & -2.28$\times$$10^{-5}$ & 5.90 \\
    Spanish: Twitter & -0.048 & 1.14$\times$$10^{-4}$ & -0.032 & 1.10$\times$$10^{-2}$ & -3.10$\times$$10^{-5}$ & 5.94 \\
    Portuguese: Google Web Crawl & -0.085 & 6.33$\times$$10^{-13}$ & -0.060 & 3.23$\times$$10^{-7}$ & -3.98$\times$$10^{-5}$ & 5.96 \\
    Portuguese: Twitter & -0.041 & 5.98$\times$$10^{-4}$ & -0.030 & 1.15$\times$$10^{-2}$ & -2.40$\times$$10^{-5}$ & 5.73 \\
    English: Google Books & -0.042 & 3.03$\times$$10^{-3}$ & -0.013 & 3.50$\times$$10^{-1}$ & -3.04$\times$$10^{-5}$ & 5.62 \\
    English: New York Times & -0.056 & 6.93$\times$$10^{-5}$ & -0.044 & 1.99$\times$$10^{-3}$ & -4.17$\times$$10^{-5}$ & 5.61 \\
    German: Google Web Crawl & -0.096 & 1.11$\times$$10^{-15}$ & -0.082 & 6.75$\times$$10^{-12}$ & -3.67$\times$$10^{-5}$ & 5.65 \\
    French: Google Web Crawl & -0.105 & 9.20$\times$$10^{-19}$ & -0.080 & 1.99$\times$$10^{-11}$ & -4.50$\times$$10^{-5}$ & 5.68 \\
    English: Twitter & -0.097 & 6.56$\times$$10^{-12}$ & -0.103 & 2.37$\times$$10^{-13}$ & -7.78$\times$$10^{-5}$ & 5.67 \\
    Indonesian: Movie subtitles & -0.039 & 1.48$\times$$10^{-3}$ & -0.063 & 2.45$\times$$10^{-7}$ & -2.04$\times$$10^{-5}$ & 5.45 \\
    German: Twitter & -0.054 & 1.47$\times$$10^{-5}$ & -0.036 & 4.02$\times$$10^{-3}$ & -2.51$\times$$10^{-5}$ & 5.58 \\
    Russian: Twitter & -0.052 & 2.38$\times$$10^{-5}$ & -0.028 & 2.42$\times$$10^{-2}$ & -2.55$\times$$10^{-5}$ & 5.52 \\
    French: Google Books & -0.043 & 6.80$\times$$10^{-4}$ & -0.030 & 1.71$\times$$10^{-2}$ & -2.31$\times$$10^{-5}$ & 5.49 \\
    German: Google Books & -0.003 & 8.12$\times$$10^{-1}$ & +0.014 & 2.74$\times$$10^{-1}$ & -1.38$\times$$10^{-6}$ & 5.45 \\
    French: Twitter & -0.049 & 6.08$\times$$10^{-5}$ & -0.023 & 6.31$\times$$10^{-2}$ & -2.54$\times$$10^{-5}$ & 5.54 \\
    Russian: Movie and TV subtitles & -0.029 & 2.36$\times$$10^{-2}$ & -0.033 & 9.17$\times$$10^{-3}$ & -1.57$\times$$10^{-5}$ & 5.43 \\
    Arabic: Movie and TV subtitles & -0.045 & 7.10$\times$$10^{-6}$ & -0.029 & 4.19$\times$$10^{-3}$ & -1.66$\times$$10^{-5}$ & 5.44 \\
    Indonesian: Twitter & -0.051 & 2.14$\times$$10^{-5}$ & -0.018 & 1.24$\times$$10^{-1}$ & -2.50$\times$$10^{-5}$ & 5.46 \\
    Korean: Twitter & -0.032 & 8.29$\times$$10^{-3}$ & -0.016 & 1.91$\times$$10^{-1}$ & -1.24$\times$$10^{-5}$ & 5.38 \\
    Russian: Google Books & +0.030 & 2.09$\times$$10^{-2}$ & +0.070 & 5.08$\times$$10^{-8}$ & +1.20$\times$$10^{-5}$ & 5.35 \\
    English: Music Lyrics & -0.073 & 2.53$\times$$10^{-7}$ & -0.081 & 1.05$\times$$10^{-8}$ & -6.12$\times$$10^{-5}$ & 5.45 \\
    Korean: Movie subtitles & -0.187 & 8.22$\times$$10^{-44}$ & -0.180 & 2.01$\times$$10^{-40}$ & -9.66$\times$$10^{-5}$ & 5.41 \\
    Chinese: Google Books & -0.067 & 1.48$\times$$10^{-11}$ & -0.050 & 5.01$\times$$10^{-7}$ & -1.72$\times$$10^{-5}$ & 5.21 \\
    \hline
    \hline
  \end{tabular}
  \caption{
    Pearson correlation coefficients and $p$-values,
    Spearman correlation coefficients and $p$-values,
    and linear fit coefficients, 
    for average word happiness $h_{\textnormal{avg}}$
    as a function of word usage frequency rank $r$.
    We use the fit is $h_{\textnormal{avg}} = \alpha r + \beta$
    for the most common 5000 words in each corpora,
    determining $\alpha$ and $\beta$ via ordinary least squares,
    and order languages by 
    the median of their average word happiness scores (descending).
    We note that stemming of words may affect these estimates.
  }
  \label{tab:mhl.havgcorr}
\end{table*}

\begin{table*}[htbp!]
  \centering
  \rowcolors{1}{verylightgrey}{white}
  \begin{tabular}{|l|c|c|c|c|c|c|}
    \hline
    \hline
    Language: Corpus & $\rho_{\rm p}$ & $p$-value & $\rho_{\rm s}$ & $p$-value & $\alpha$ & $\beta$ \\
    \hline
    Portuguese: Twitter & +0.090 & 2.55$\times$$10^{-14}$ & +0.095 & 1.28$\times$$10^{-15}$ & 1.19$\times$$10^{-5}$ & 1.29 \\
    Spanish: Twitter & +0.097 & 8.45$\times$$10^{-15}$ & +0.104 & 5.92$\times$$10^{-17}$ & 1.47$\times$$10^{-5}$ & 1.26 \\
    English: Music Lyrics & +0.129 & 4.87$\times$$10^{-20}$ & +0.134 & 1.63$\times$$10^{-21}$ & 2.76$\times$$10^{-5}$ & 1.33 \\
    English: Twitter & +0.007 & 6.26$\times$$10^{-1}$ & +0.012 & 4.11$\times$$10^{-1}$ & 1.47$\times$$10^{-6}$ & 1.35 \\
    English: New York Times & +0.050 & 4.56$\times$$10^{-4}$ & +0.044 & 1.91$\times$$10^{-3}$ & 9.34$\times$$10^{-6}$ & 1.32 \\
    Arabic: Movie and TV subtitles & +0.101 & 7.13$\times$$10^{-24}$ & +0.101 & 3.41$\times$$10^{-24}$ & 9.41$\times$$10^{-6}$ & 1.01 \\
    English: Google Books & +0.180 & 1.68$\times$$10^{-37}$ & +0.176 & 4.96$\times$$10^{-36}$ & 3.36$\times$$10^{-5}$ & 1.27 \\
    Spanish: Google Books & +0.066 & 1.23$\times$$10^{-7}$ & +0.062 & 6.53$\times$$10^{-7}$ & 9.17$\times$$10^{-6}$ & 1.26 \\
    Indonesian: Movie subtitles & +0.026 & 3.43$\times$$10^{-2}$ & +0.027 & 2.81$\times$$10^{-2}$ & 2.87$\times$$10^{-6}$ & 1.12 \\
    Russian: Movie and TV subtitles & +0.083 & 7.60$\times$$10^{-11}$ & +0.075 & 3.28$\times$$10^{-9}$ & 1.06$\times$$10^{-5}$ & 0.89 \\
    French: Twitter & +0.072 & 4.77$\times$$10^{-9}$ & +0.076 & 8.94$\times$$10^{-10}$ & 1.07$\times$$10^{-5}$ & 1.05 \\
    Indonesian: Twitter & +0.072 & 1.17$\times$$10^{-9}$ & +0.072 & 1.73$\times$$10^{-9}$ & 8.16$\times$$10^{-6}$ & 1.12 \\
    French: Google Books & +0.090 & 1.02$\times$$10^{-12}$ & +0.085 & 1.67$\times$$10^{-11}$ & 1.25$\times$$10^{-5}$ & 1.02 \\
    Russian: Twitter & +0.055 & 6.83$\times$$10^{-6}$ & +0.053 & 1.67$\times$$10^{-5}$ & 7.39$\times$$10^{-6}$ & 0.91 \\
    Spanish: Google Web Crawl & +0.119 & 4.45$\times$$10^{-24}$ & +0.106 & 2.60$\times$$10^{-19}$ & 1.45$\times$$10^{-5}$ & 1.23 \\
    Portuguese: Google Web Crawl & +0.093 & 4.06$\times$$10^{-15}$ & +0.083 & 2.91$\times$$10^{-12}$ & 1.07$\times$$10^{-5}$ & 1.26 \\
    German: Twitter & +0.051 & 4.45$\times$$10^{-5}$ & +0.050 & 5.15$\times$$10^{-5}$ & 7.39$\times$$10^{-6}$ & 1.15 \\
    French: Google Web Crawl & +0.104 & 2.12$\times$$10^{-18}$ & +0.088 & 9.64$\times$$10^{-14}$ & 1.27$\times$$10^{-5}$ & 1.01 \\
    Korean: Movie subtitles & +0.171 & 1.39$\times$$10^{-36}$ & +0.185 & 8.85$\times$$10^{-43}$ & 2.58$\times$$10^{-5}$ & 0.88 \\
    German: Google Books & +0.157 & 6.06$\times$$10^{-35}$ & +0.162 & 4.96$\times$$10^{-37}$ & 2.17$\times$$10^{-5}$ & 1.03 \\
    Korean: Twitter & +0.056 & 4.07$\times$$10^{-6}$ & +0.062 & 4.25$\times$$10^{-7}$ & 6.98$\times$$10^{-6}$ & 0.93 \\
    German: Google Web Crawl & +0.099 & 2.05$\times$$10^{-16}$ & +0.085 & 1.18$\times$$10^{-12}$ & 1.20$\times$$10^{-5}$ & 1.07 \\
    Chinese: Google Books & +0.099 & 3.07$\times$$10^{-23}$ & +0.097 & 3.81$\times$$10^{-22}$ & 8.70$\times$$10^{-6}$ & 1.16 \\
    Russian: Google Books & +0.187 & 5.15$\times$$10^{-48}$ & +0.177 & 2.24$\times$$10^{-43}$ & 2.28$\times$$10^{-5}$ & 0.81 \\
    \hline
    \hline
  \end{tabular}
  \caption{
    Pearson correlation coefficients and $p$-values,
    Spearman correlation coefficients and $p$-values,
    and linear fit coefficients
    for standard deviation of word happiness $\hstdfn$
    as a function of word usage frequency rank $r$.
    We consider the fit is $h_{\textnormal{std}} = \alpha r + \beta$
    for the most common 5000 words in each corpora,
    determining $\alpha$ and $\beta$ via ordinary least squares,
    and order corpora according to their emotional variance (descending).
  }
  \label{tab:mhl.hstdcorr}
\end{table*}

\begin{figure*}
  \centering
  \includegraphics[width=\textwidth]{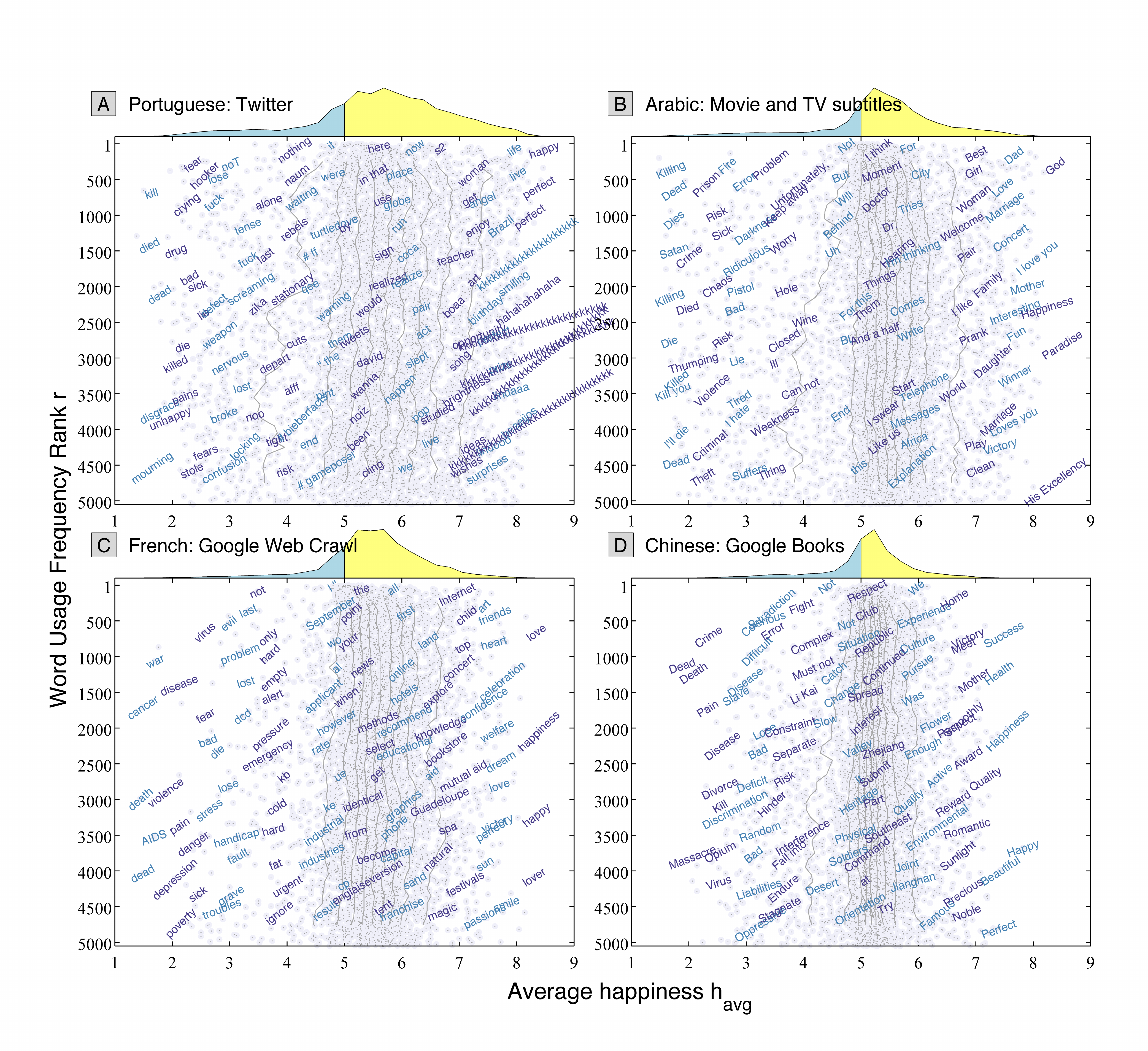}
  \caption{
    Reproduction of Fig.~\ref{fig:mlhap.jellyfish} in the main text
    with words directly translated into English
    using Google Translate.
    See the caption of Fig.~\ref{fig:mlhap.jellyfish} for details.
  }
  \label{fig:mlhap.jellyfish_translated}
\end{figure*}

\begin{figure*}[tbp!]
  \centering
 \includegraphics[width=\textwidth]{fig009_fighappinessdist_jellyfish_words_havg_multilanguage001_combined001.pdf}
  \caption{
    Jellyfish plots showing how average word happiness distribution is
    strongly invariant with respect to word rank for corpora ranked
    1--6 according to median word happiness.
    See the caption of Fig.~\ref{fig:mlhap.jellyfish} in the main text for details.
  }
  \label{fig:mlhap.happinessdist_jellyfish_words_havg_multilanguage001_table1}
\end{figure*}

\begin{figure*}[tbp!]
  \centering
 \includegraphics[width=\textwidth]{fig010_fighappinessdist_jellyfish_words_havg_multilanguage001_combined002.pdf}
  \caption{
    Jellyfish plots showing how average word happiness distribution is
    strongly invariant with respect to word rank for corpora ranked
    7--12 according to median word happiness.
    See the caption of Fig.~\ref{fig:mlhap.jellyfish} in the main text for details.
  }
  \label{fig:mlhap.happinessdist_jellyfish_words_havg_multilanguage001_table2}
\end{figure*}

\begin{figure*}[tbp!]
  \centering
 \includegraphics[width=\textwidth]{fig011_fighappinessdist_jellyfish_words_havg_multilanguage001_combined003.pdf}
  \caption{
    Jellyfish plots showing how average word happiness distribution is
    strongly invariant with respect to word rank for corpora ranked
    13--18 according to median word happiness.
    See the caption of Fig.~\ref{fig:mlhap.jellyfish} in the main text for details.
  }
  \label{fig:mlhap.happinessdist_jellyfish_words_havg_multilanguage001_table3}
\end{figure*}

\begin{figure*}[tbp!]
  \centering
 \includegraphics[width=\textwidth]{fig012_fighappinessdist_jellyfish_words_havg_multilanguage001_combined004.pdf}  
  \caption{
    Jellyfish plots showing how average word happiness distribution is
    strongly invariant with respect to word rank for corpora ranked
    19--24 according to median word happiness.
    See the caption of Fig.~\ref{fig:mlhap.jellyfish} in the main text for details.
  }
  \label{fig:mlhap.happinessdist_jellyfish_words_havg_multilanguage001_table4}
\end{figure*}

\begin{figure*}[tbp!]
  \centering
 \includegraphics[width=\textwidth]{fig013_fighappinessdist_jellyfish_words_havg_multilanguage001_combined005.pdf}  
  \caption{
    Jellyfish plots showing how standard deviation of word happiness
    behaves with respect to word rank for corpora ranked
    1--6 according to median word happiness.
    See the caption of Fig.~\ref{fig:mlhap.jellyfish} in the main text for details.
  }
  \label{fig:mlhap.happinessdist_jellyfish_words_hstd_multilanguage001_table1}
\end{figure*}

\begin{figure*}[tbp!]
  \centering
 \includegraphics[width=\textwidth]{fig014_fighappinessdist_jellyfish_words_havg_multilanguage001_combined006.pdf}  
  \caption{
    Jellyfish plots showing how standard deviation of word happiness
    behaves with respect to word rank for corpora ranked
    7--12 according to median word happiness.
    See the caption of Fig.~\ref{fig:mlhap.jellyfish} in the main text for details.
  }
  \label{fig:mlhap.happinessdist_jellyfish_words_hstd_multilanguage001_table2}
\end{figure*}

\begin{figure*}[tbp!]
  \centering
 \includegraphics[width=\textwidth]{fig015_fighappinessdist_jellyfish_words_havg_multilanguage001_combined007.pdf}  
  \caption{
    Jellyfish plots showing how standard deviation of word happiness
    behaves with respect to word rank for corpora ranked
    13--18 according to median word happiness.
    See the caption of Fig.~\ref{fig:mlhap.jellyfish} in the main text for details.
  }
  \label{fig:mlhap.happinessdist_jellyfish_words_hstd_multilanguage001_table3}
\end{figure*}

\begin{figure*}[tbp!]
  \centering
 \includegraphics[width=\textwidth]{fig016_fighappinessdist_jellyfish_words_havg_multilanguage001_combined008.pdf}  
  \caption{
    Jellyfish plots showing how standard deviation of word happiness
    behaves with respect to word rank for corpora ranked
    19--24 according to median word happiness.
    See the caption of Fig.~\ref{fig:mlhap.jellyfish} in the main text for details.
  }
  \label{fig:mlhap.happinessdist_jellyfish_words_hstd_multilanguage001_table4}
\end{figure*}

\begin{figure*}[tbp!]
  \centering
 \includegraphics[width=\textwidth]{fig017_fighappinessdist_jellyfish_words_havg_multilanguage001_combined009.pdf}  
  \caption{
    English-translated Jellyfish plots showing how average word happiness distribution is
    strongly invariant with respect to word rank for corpora ranked
    1--6 according to median word happiness.
    See the caption of Fig.~\ref{fig:mlhap.jellyfish} in the main text for details.
  }
  \label{fig:mlhap.happinessdist_jellyfish_words_havg_multilanguage002_table1}
\end{figure*}

\begin{figure*}[tbp!]
  \centering
  \includegraphics[width=\textwidth]{fig018_fighappinessdist_jellyfish_words_havg_multilanguage001_combined010.pdf}
  \caption{
    English-translated Jellyfish plots showing how average word happiness distribution is
    strongly invariant with respect to word rank for corpora ranked
    7--12 according to median word happiness.
    See the caption of Fig.~\ref{fig:mlhap.jellyfish} in the main text for details.
  }
  \label{fig:mlhap.happinessdist_jellyfish_words_havg_multilanguage002_table2}
\end{figure*}

\begin{figure*}[tbp!]
  \centering
  \includegraphics[width=\textwidth]{fig019_fighappinessdist_jellyfish_words_havg_multilanguage001_combined011.pdf}
  \caption{
    English-translated Jellyfish plots showing how average word happiness distribution is
    strongly invariant with respect to word rank for corpora ranked
    13--18 according to median word happiness.
    See the caption of Fig.~\ref{fig:mlhap.jellyfish} in the main text for details.
  }
  \label{fig:mlhap.happinessdist_jellyfish_words_havg_multilanguage002_table3}
\end{figure*}

\begin{figure*}[tbp!]
  \centering
  \includegraphics[width=\textwidth]{fig020_fighappinessdist_jellyfish_words_havg_multilanguage001_combined012.pdf}
  \caption{
    English-translated Jellyfish plots showing how average word happiness distribution is
    strongly invariant with respect to word rank for corpora ranked
    19--24 according to median word happiness.
    See the caption of Fig.~\ref{fig:mlhap.jellyfish} in the main text for details.
  }
  \label{fig:mlhap.happinessdist_jellyfish_words_havg_multilanguage002_table4}
\end{figure*}

\begin{figure*}[tbp!]
  \centering
  \includegraphics[width=\textwidth]{fig021_fighappinessdist_jellyfish_words_havg_multilanguage001_combined013.pdf}
  \caption{
    English-translated Jellyfish plots showing how standard deviation of word happiness
    behaves with respect to word rank for corpora ranked
    1--6 according to median word happiness.
    See the caption of Fig.~\ref{fig:mlhap.jellyfish} in the main text for details.
  }
  \label{fig:mlhap.happinessdist_jellyfish_words_hstd_multilanguage002_table1}
\end{figure*}

\begin{figure*}[tbp!]
  \centering
  \includegraphics[width=\textwidth]{fig022_fighappinessdist_jellyfish_words_havg_multilanguage001_combined014.pdf}
  \caption{
    English-translated Jellyfish plots showing how standard deviation of word happiness
    behaves with respect to word rank for corpora ranked
    7--12 according to median word happiness.
    See the caption of Fig.~\ref{fig:mlhap.jellyfish} in the main text for details.
  }
  \label{fig:mlhap.happinessdist_jellyfish_words_hstd_multilanguage002_table2}
\end{figure*}

\begin{figure*}[tbp!]
  \centering
  \includegraphics[width=\textwidth]{fig023_fighappinessdist_jellyfish_words_havg_multilanguage001_combined015.pdf}
  \caption{
    English-translated Jellyfish plots showing how standard deviation of word happiness
    behaves with respect to word rank for corpora ranked
    13--18 according to median word happiness.
    See the caption of Fig.~\ref{fig:mlhap.jellyfish} in the main text for details.
  }
  \label{fig:mlhap.happinessdist_jellyfish_words_hstd_multilanguage002_table3}
\end{figure*}

\begin{figure*}[tbp!]
  \centering
  \includegraphics[width=\textwidth]{fig024_fighappinessdist_jellyfish_words_havg_multilanguage001_combined016.pdf}
  \caption{
    English-translated Jellyfish plots showing how standard deviation of word happiness
    behaves with respect to word rank for corpora ranked
    19--24 according to median word happiness.
    See the caption of Fig.~\ref{fig:mlhap.jellyfish} in the main text for details.
  }
  \label{fig:mlhap.happinessdist_jellyfish_words_hstd_multilanguage002_table4}
\end{figure*}

\begin{figure*}
  \centering
  \includegraphics[width=\textwidth]{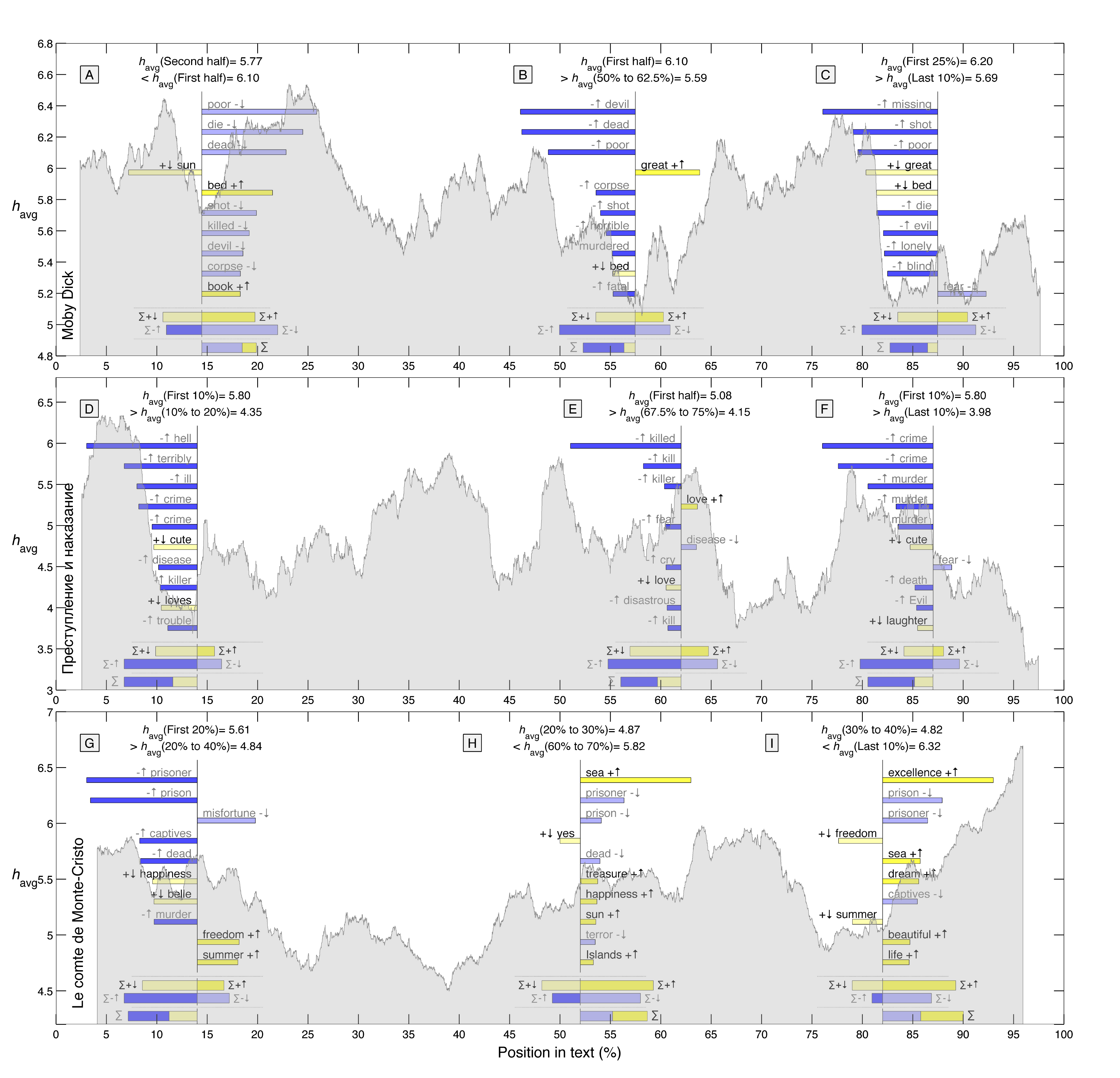}
  \caption{
    Fig.~\ref{fig:mlhap.measurementexamples} from the main text
    with Russian and French translated into English.
  }
  \label{fig:mlhap.measurementexamples_translated}
\end{figure*}

\clearpage

\section*{Explanation of Word Shifts}
\label{sec:mlhap.wordshiftexplanation}
\label{page:mhl.startwordshiftdescription}

In this section, we explain our word shifts in detail,
both the abbreviated ones included in
Figs.~\ref{fig:mlhap.measurementexamples}
and~\ref{fig:mlhap.measurementexamples_translated},
and the more sophisticated, complementary word shifts which
follow in this supplementary section.
We expand upon the approach described in~\cite{dodds2009b}
and~\cite{dodds2011e} to rank and visualize how words contribute
to this overall upward shift in happiness.

Shown below is the third inset word shift
used in Fig~\ref{fig:mlhap.measurementexamples}
for the Count of Monte Cristo,
a comparison of words found in the last 10\% of the book ($\Tcomp$, $\havgfn$ = 6.32)
relative to those used between 30\% and 40\% ($\Tref$, $\havgfn$ = 4.82).
For this particular measurement, we employed the `word lens' which excluded words 
with $3 < \havgfn <7$.

\begin{myframe}
  \begin{center}
    \includegraphics[width=0.8\columnwidth]{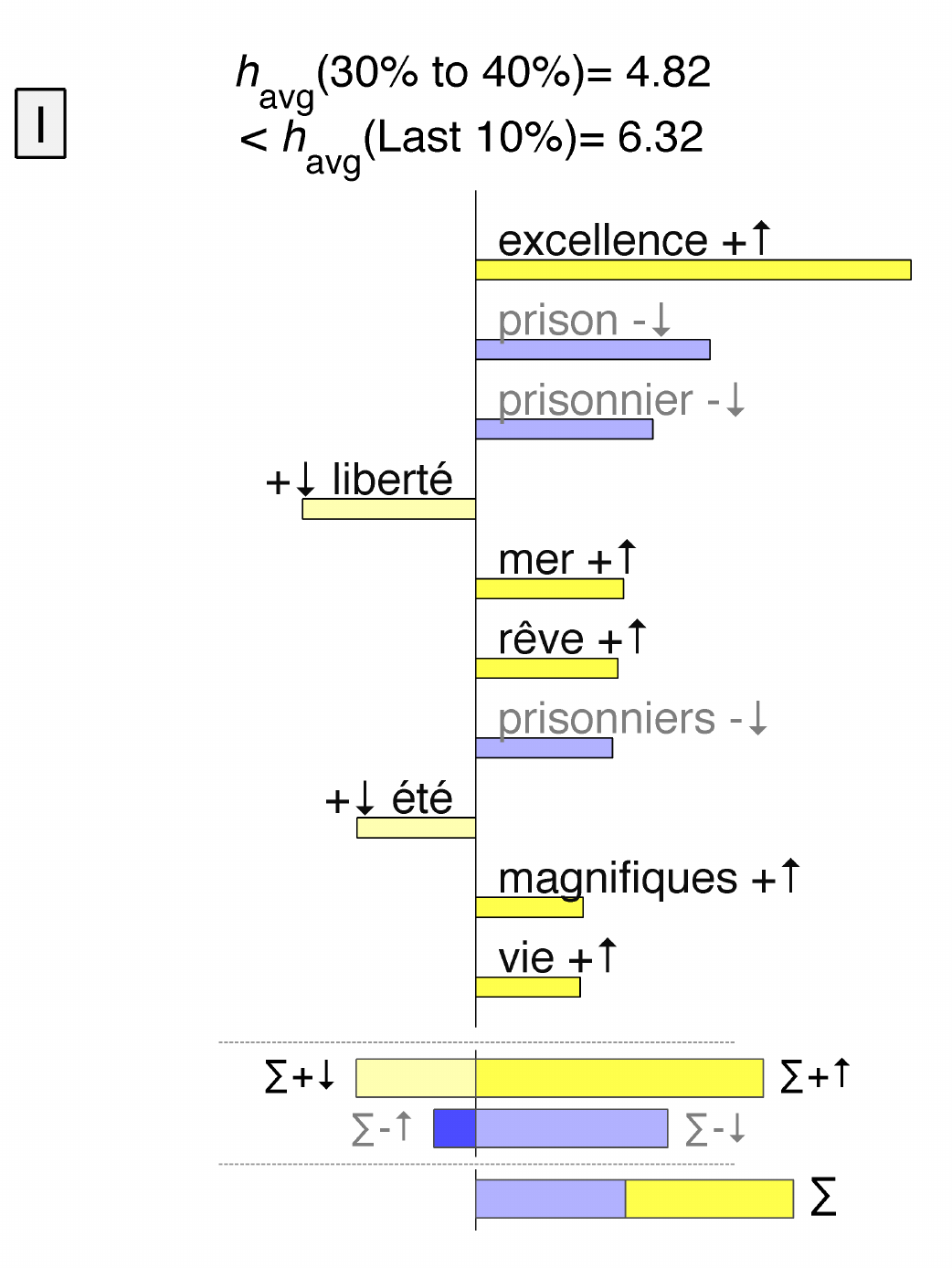}
  \end{center}
\end{myframe}

We will use the following probability notation for 
the normalized frequency of a given word $w$ in a text $T$:
\begin{equation}
  \Probof{w|T;L}
  =
  \frac{
    f(w|T;L)
  }
  {
    \sum_{w' \in L}
    f(w'|T;L)
  }
  \label{eq:mlh.probfreqhapp},
\end{equation}
where $f(w|T;L)$ is the frequency of word $w$ in $T$
with word lens $L$ applied~\cite{dodds2009b}.
(For the example word shift above, we have $L = \{[1, 3],[7, 9]\}$.)
We then estimate the happiness score of any text $T$ as
\begin{equation}
  \havg{T;L} = 
  \sum_{w \in L}
  \havg{w}
  \Probof{w|T;L},
  \label{eq:mlh.texthappiness}
\end{equation}
where $\havg{w}$ is the average happiness score of a word
as determined by our survey.

We can now express the happiness difference 
between two texts as follows:
\begin{eqnarray}
  \lefteqn{
    \havg{\Tcomp;L}
    - 
    \havg{\Tref;L}} 
  \nonumber \\
  & = &
  \sum_{w \in L}
  \havg{w}
  \Probof{w|\Tcomp;L}
  -
  \sum_{w \in L}
  \havg{w}
  \Probof{w|\Tref;L} 
  \nonumber \\
  & = &
  \sum_{w \in L}
  \havg{w}
  \left[
  \Probof{w|\Tcomp;L}
  -
  \Probof{w|\Tref;L} 
  \right]
  \nonumber \\
  & = &
  \sum_{w \in L}
  \left[
    \havg{w} - \havg{\Tref;L}
  \right]
  \left[
  \Probof{w|\Tcomp;L}
  -
  \Probof{w|\Tref;L} 
  \right],
  \nonumber \\
  \label{eq:mlh.deltah}
\end{eqnarray}
where we have introduced $\havg{\Tref;L}$ as 
base reference for the average happiness of a word
by noting that 
\begin{gather}
\sum_{w \in L}
\havg{\Tref;L}
\left[
  \Probof{w|\Tcomp;L}
  -
  \Probof{w|\Tref;L} 
\right]
\nonumber \\
=
\havg{\Tref;L}
\sum_{w \in L}
\left[
  \Probof{w|\Tcomp;L}
  -
  \Probof{w|\Tref;L} 
\right]
\nonumber \\
=
\havg{\Tref;L}
\left[
  1
  -
  1
\right]
=
0.
\end{gather}

We can now see the change in average happiness between
a reference and comparison text as depending on how these two quantities
behave for each word:
\begin{equation}
  \deltahw
  =
  \left[
    \havg{w} - \havg{\Tref;L}
  \right]
  \label{eq:mlh.hapshift}
\end{equation}
and
\begin{equation}
  \deltapw
  =
  \left[
    \Probof{w|\Tcomp;L}
    -
    \Probof{w|\Tref;L} 
  \right].
  \label{eq:mlh.probshift}
\end{equation}
Words can contribute to or work against a shift in average
happiness in four possible ways which we encode with
symbols and colors:
\begin{itemize}
\item 
  \textbf{$\deltahw > 0$, $\deltapw > 0$:}
  Words that are more positive than the reference
  text's overall average and are used more in the comparison text
  ($+$$\uparrow$, strong yellow).
\item 
  \textbf{$\deltahw < 0$, $\deltapw < 0$:}
  Words that are less positive than the reference
  text's overall average but are used less in the comparison text
  ($-$$\downarrow$, pale blue).
\item 
  \textbf{$\deltahw > 0$, $\deltapw < 0$:}
  Words that are more positive than the reference
  text's overall average but are used less in the comparison text
  ($+$$\downarrow$, pale yellow).
\item 
  \textbf{$\deltahw < 0$, $\deltapw > 0$:}
  Words that are more positive than the reference
  text's overall average and are used more in the comparison text
  ($-$$\uparrow$, strong blue).
\end{itemize}
Regardless of usage changes, 
yellow indicates a relatively positive word, blue a negative one.
The stronger colors indicate words with the most simple impact:
relatively positive or negative words being used more overall.

We order words by the absolute value of their contribution to or
against the overall shift,
and normalize them as percentages.

\subsection*{Simple Word Shifts}

For simple inset word shifts, we show the 10 top
words in terms of their absolute contribution to the shift.

Returning to the inset word shift above, we see that an
increase in the abundance of relatively positive words
`excellence'
`mer'
and 
`r\^{e}ve'
($+$$\uparrow$, strong yellow)
as well as a decrease in the relatively negative words
`prison'
and
`prisonnier'
($-$$\downarrow$, pale blue)
most strongly contribute 
to the increase in positivity.
Some words go against this trend, and in the abbreviated word shift
we see less usage of relatively positive words
`libert\'{e}'
and
`\'{e}t\'{e}'
($+$$\downarrow$, pale yellow).

The normalized sum total of each of the four categories of 
words is shown in the summary bars at the bottom of the
word shift.  For example, $\Sigma$$+$$\uparrow$ represents
the total shift due to all relatively positive words
that are more prevalent in the comparison text.
The smallest contribution comes from relatively negative
words being used more 
($-$$\uparrow$, strong blue).

The bottom bar with $\Sigma$ shows the overall shift
with a breakdown of how relatively positive and negative 
words separately contribute.  For the Count of Monte Cristo
example, we observe an overall use of relatively positive words and a drop
in the use of relatively negative ones (strong yellow and pale blue).

\subsection*{Full Word Shifts}

\begin{figure*}[tp!]
  \centering
  \includegraphics[height=0.9\textheight]{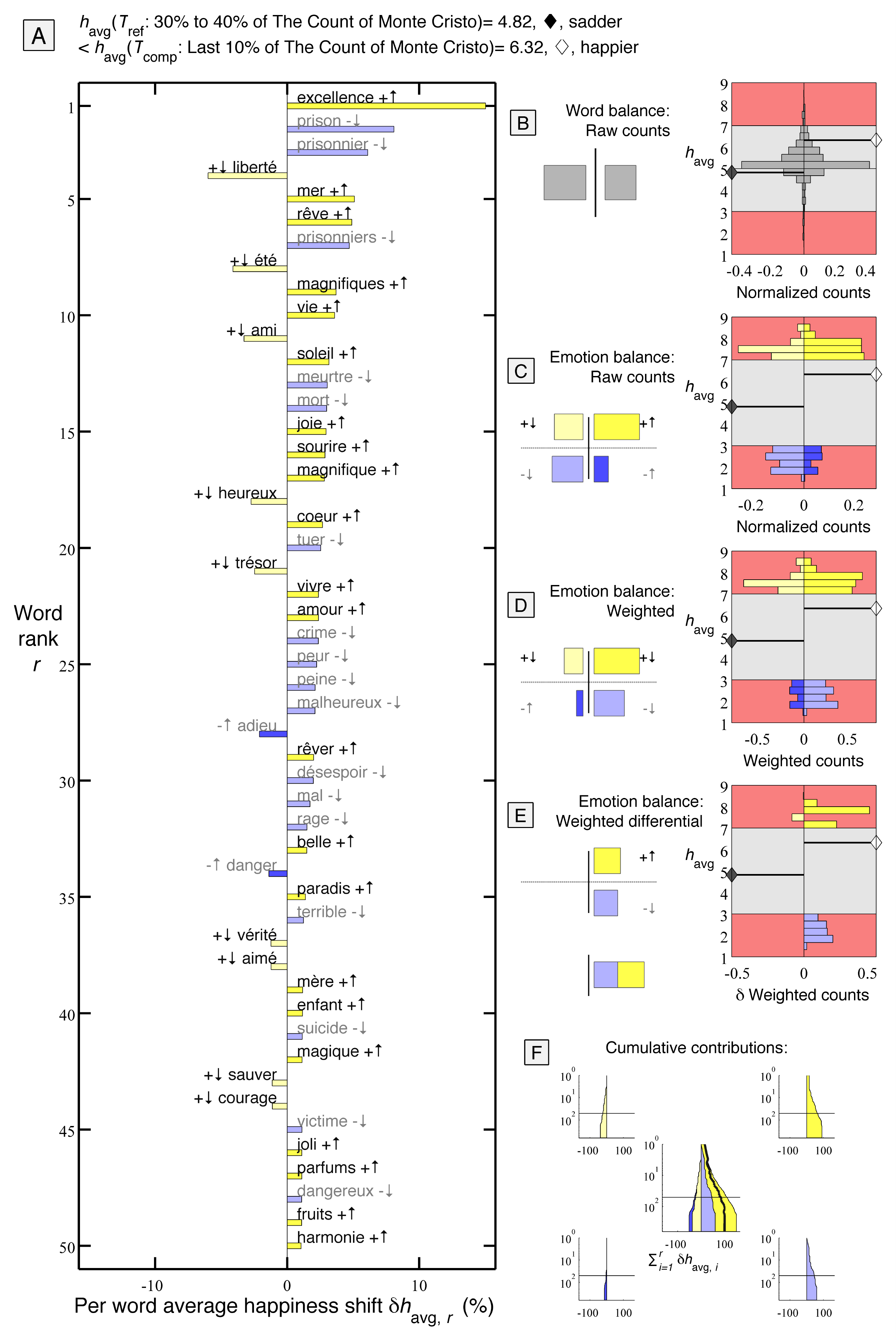}
  \caption{
    Detailed version of the 
    third word shift 
    for the Count of Monte Cristo
    in Fig.~\ref{fig:mlhap.measurementexamples}.
    See pp.~\pageref{page:mhl.startwordshiftdescription}--\pageref{page:mhl.endwordshiftdescription} for a full explanation.
   }
  \label{fig:universal_wordshift500_figuniversal_wordshift_hli_count_of_monte_cristo003}
\end{figure*}

We turn now to explaining the sophisticated word shifts we include
at the end of this document.  
We break down the full word shift corresponding to the simple
one we have just addressed for the Count of Monte Cristo, 
Fig.~\ref{fig:universal_wordshift500_figuniversal_wordshift_hli_count_of_monte_cristo003}.

First, each word shift has a summary at the top:

\medskip

\begin{myframe}
\includegraphics[width=\columnwidth]{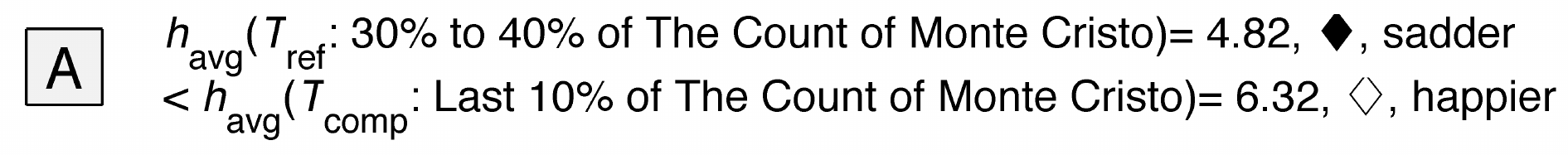}
\end{myframe}

which describes both the reference and summary text,
gives their average happiness scores,
shows which is happier through an inequality,
and functions as a legend showing that 
average happiness will be marked on graphs with diamonds (filled for 
the reference text, unfilled for the comparison one).

We note that if two texts are equal in happiness two two decimal
places, the word shift will show them as approximately
the same.  The word shift is still very much informative
as word usage will most likely have be different between
any two large-scale texts.

Below the summary and taking up the left column of each
figure, is the word shift itself for the first 50 words,
ordered by contribution rank:

\begin{myframe}
\begin{center}
\includegraphics[width=\columnwidth]{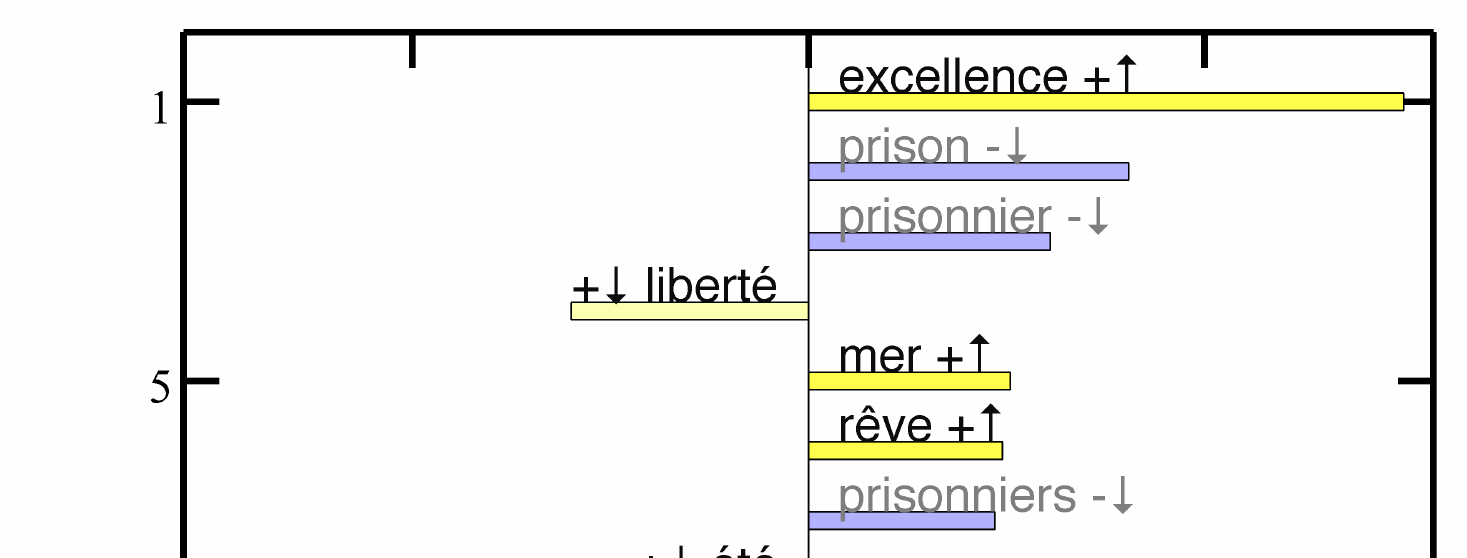}\\
\medskip
\qquad\quad\mbox{}\vdots\hfill\vdots\hfill\vdots\quad\\
\medskip
\includegraphics[width=\columnwidth]{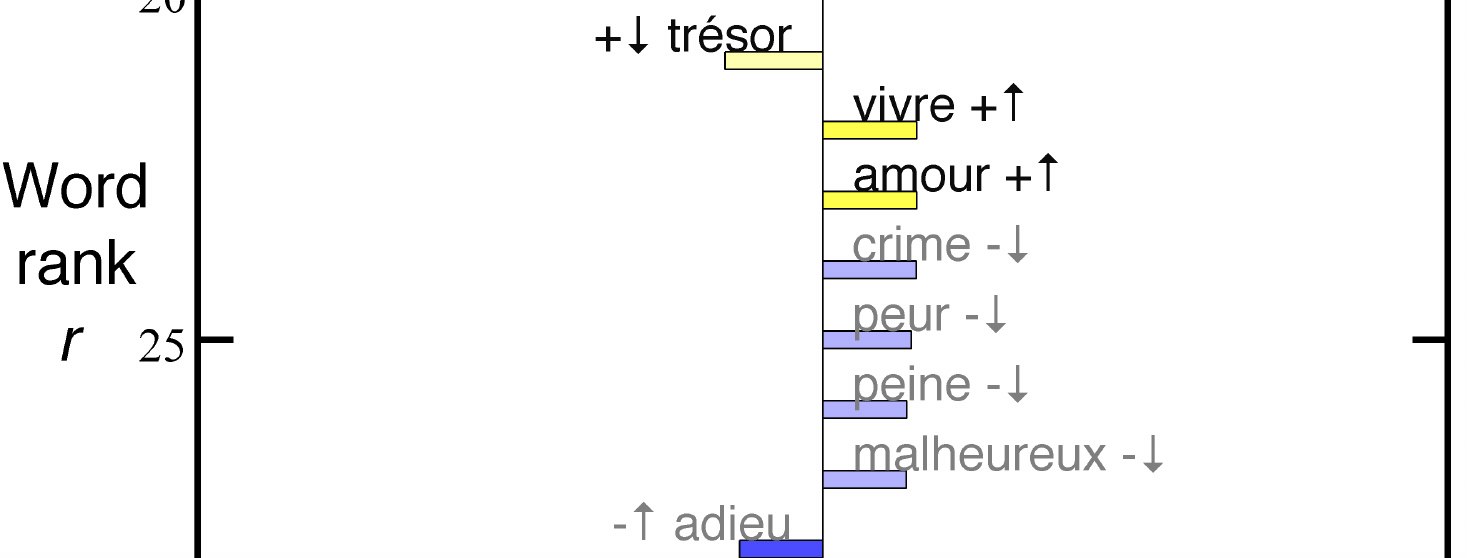}\\
\medskip
\qquad\quad\mbox{}\vdots\hfill\vdots\hfill\vdots\ \mbox{}\\
\medskip
\includegraphics[width=\columnwidth]{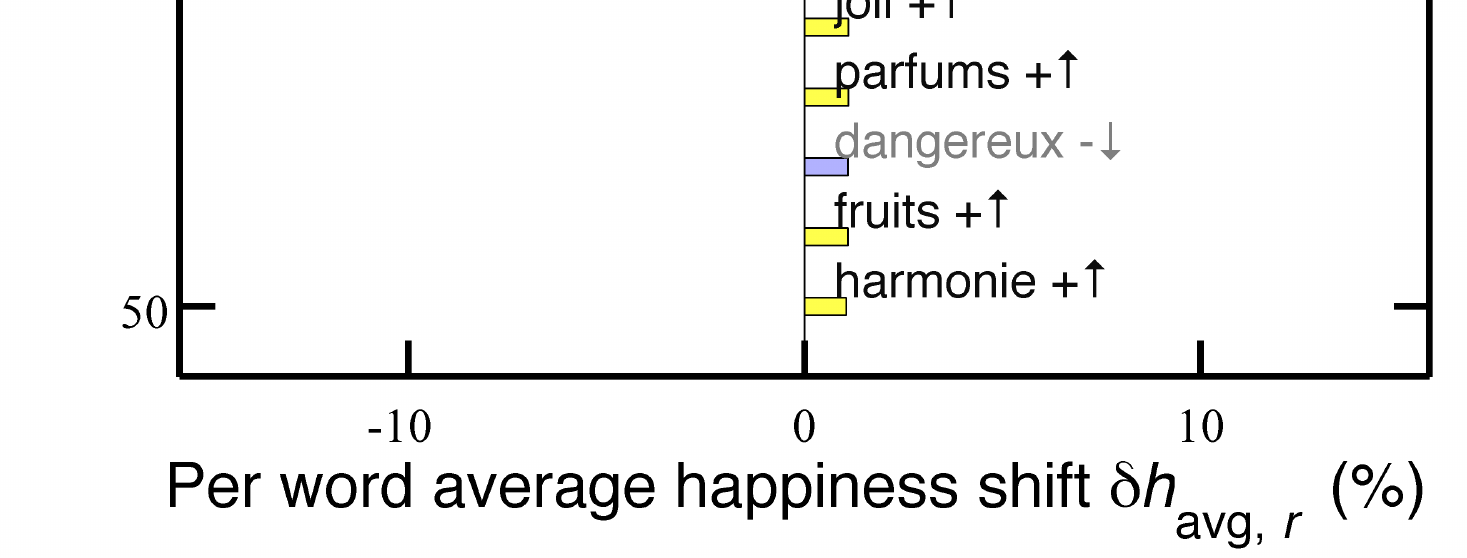}
\end{center}
\end{myframe}

The right column of each figure contains a series
of summary and histogram graphics that show how the underlying 
word distributions for each text give rise to
the overall shift.  
In all cases, and in the manner of the word shift, data for 
the reference text is on the left, the comparison is on the right.
In the histograms, we indicate the lens with a pale red for inclusion, light gray for
exclusion.  
We mark average happiness for each text by black
and unfilled diamonds.

First in plot B, we have the bare frequency distributions 
for each text.  The left hand summary compares
the sizes of the two texts (the reference is larger
in this case), while the histogram gives a detailed
view of how each text's words are distributed
according to average happiness.

\medskip

\begin{myframe}
  \includegraphics[width=\columnwidth]{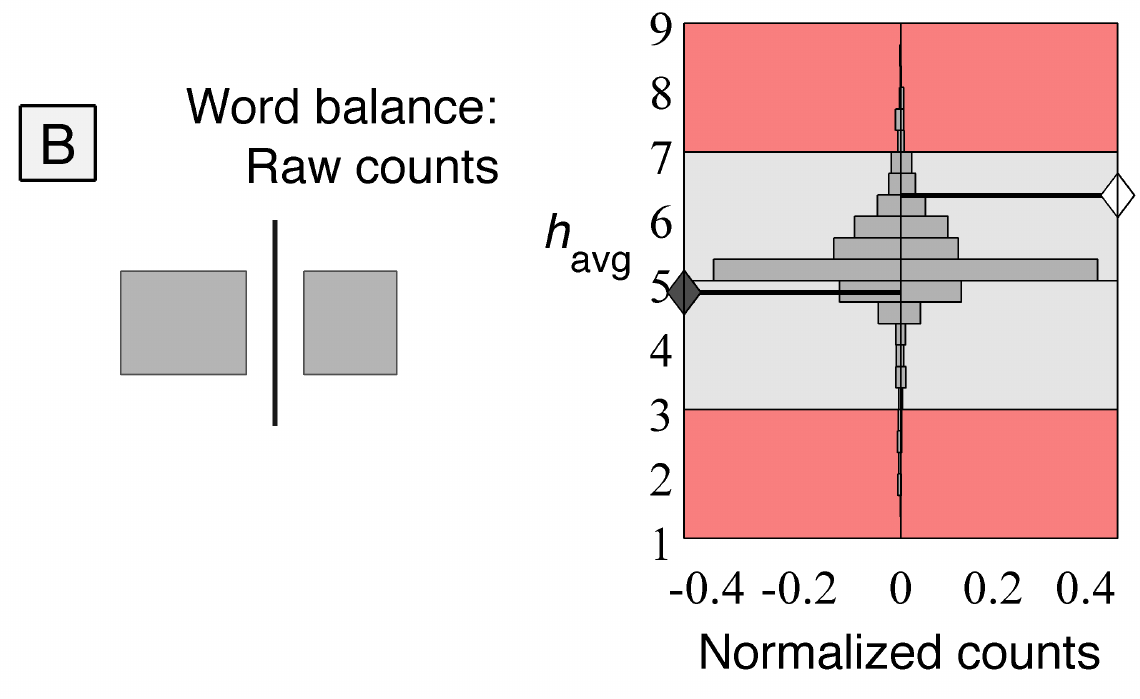}
\end{myframe}

In plot C, we then apply the lens and renormalize.
We can now also use our colors to show the relative
positivity or negativity of words.  
Note that 
the strong yellow and blue appear on the 
side of comparison 
text, as these words are being used more relative
to the reference text, and we are still considering
normalized word counts only.
The plot on the left shows the sum of the four
kinds of counts.  We can see that relatively positive words
are dominating in terms of pure counts at this stage of the computation.

\medskip

\begin{myframe}
  \includegraphics[width=\columnwidth]{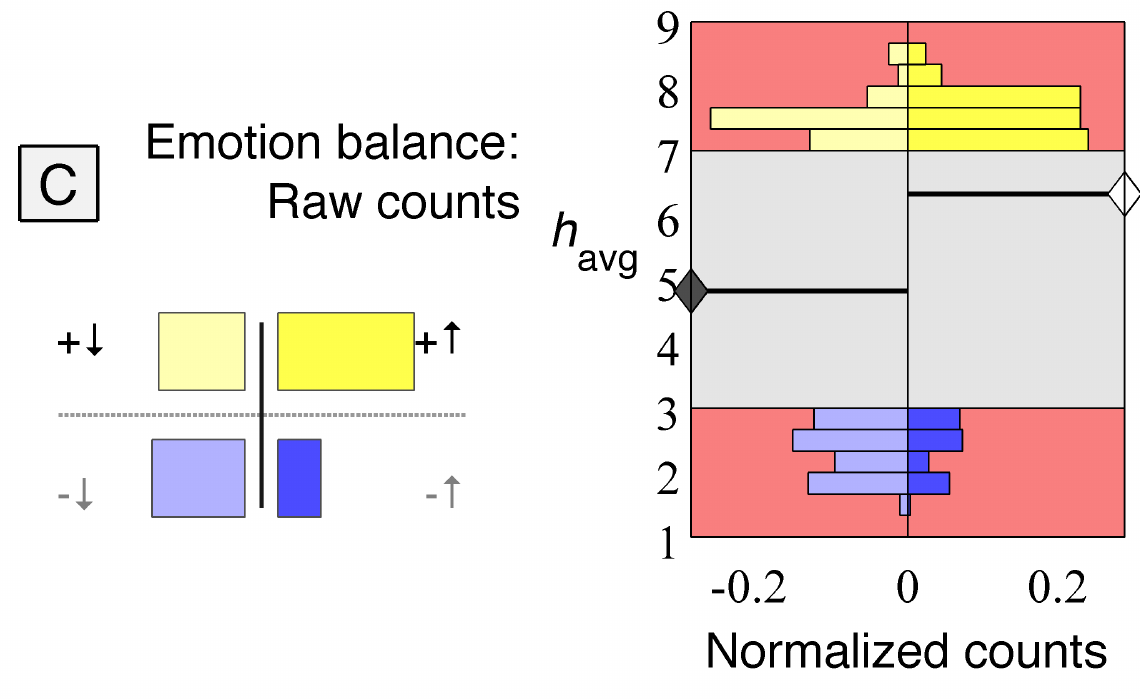}
\end{myframe}

We move to plot D, where we weight words by their
emotional distance from the reference text, $\deltahw$.
We note that in this particular example, the reference
text's average happiness is near neutral ($havgfn$ = 5),
so the shapes of histograms do not change greatly.
Also, since $\deltahw$ is negative, the colors for
the relatively negative words swap from left to right.
More frequently used negative words, for example, 
drag the comparison text down (strong blue) and
must switch toward favoring the reference text.

\medskip

\begin{myframe}
  \includegraphics[width=\columnwidth]{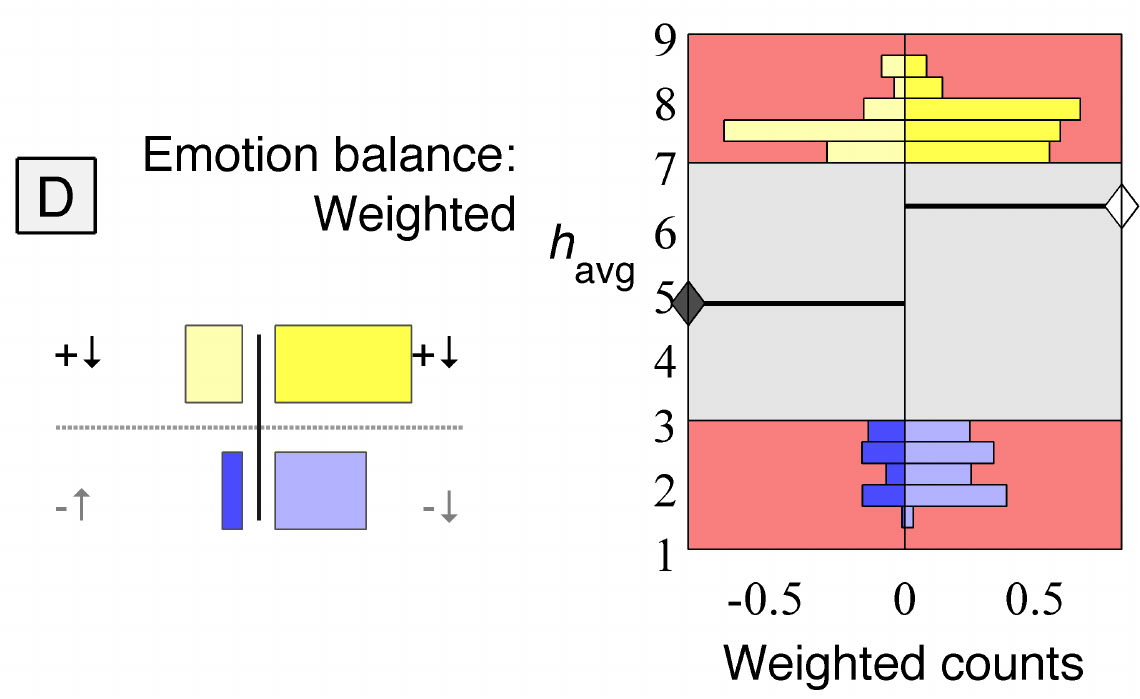}
\end{myframe}

In plot E, we incorporate the differences in word
usage, $\deltapw$.  The histogram shows the result
binned by average happiness, and in this case we
see that the comparison text is generally happier across
the negativity-positivity scale.
The summary plot shows both the sums of relatively
positive and negative words, and the overall differential.
These three bars match those at the bottom of the 
corresponding simple word shift.

\medskip

\begin{myframe}
  \includegraphics[width=\columnwidth]{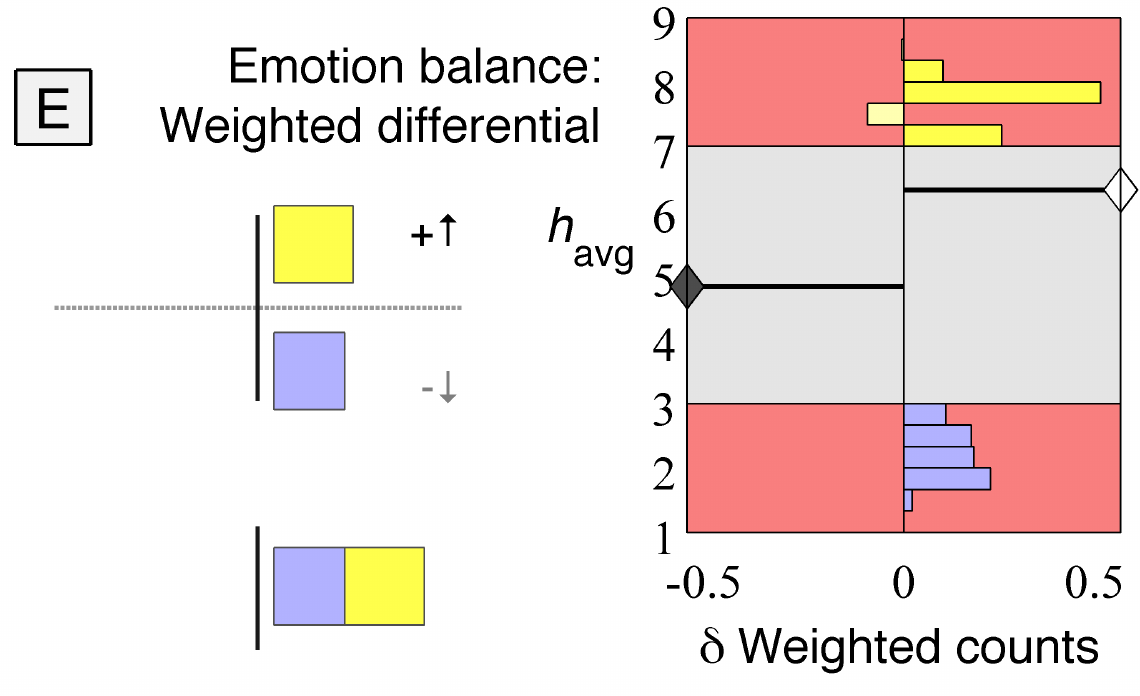}
\end{myframe}

Finally, we show how the four categories of words combine
as we sum their contributions up in descending order of absolute
contribution to or against the overall happiness shift.
The four outer plots below show the growth for each
kind of word separately, and their end points match
the bar lengths in Plot D above.  
The central plot shows how all four contribute together
with the black line showing the overall sum.
In this example, the shift is positive, and all the sum
of all contributions gives +100\%.
The horizontal line in all five plots indicates
a word rank of 50, to match the extent of Figure's word shift.

\medskip

\begin{myframe}
  \includegraphics[width=\columnwidth]{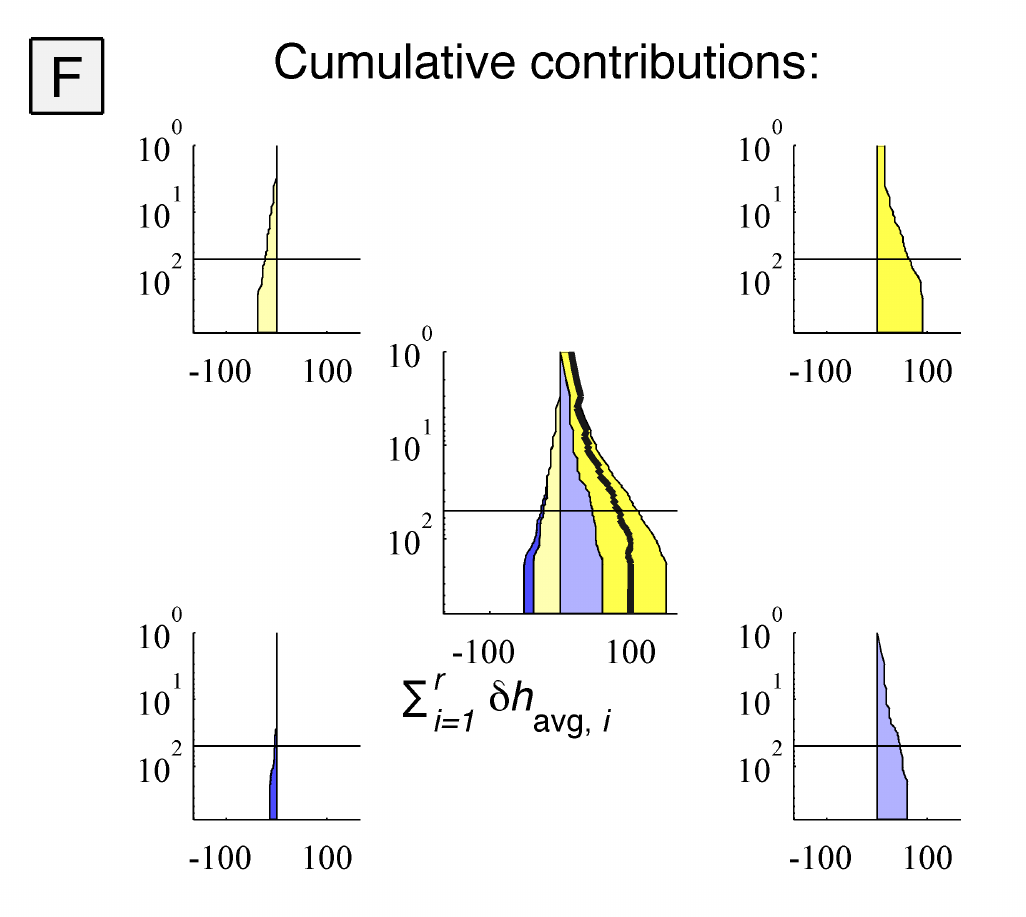}
\end{myframe}

In the remaining pages of this document, we provide
full word shifts matching the simple ones included in
Figs.~\ref{fig:mlhap.measurementexamples}
and~\ref{fig:mlhap.measurementexamples_translated}.

\label{page:mhl.endwordshiftdescription}

\clearpage

\begin{figure*}[tp!]
  \centering
  \includegraphics[height=0.9\textheight]{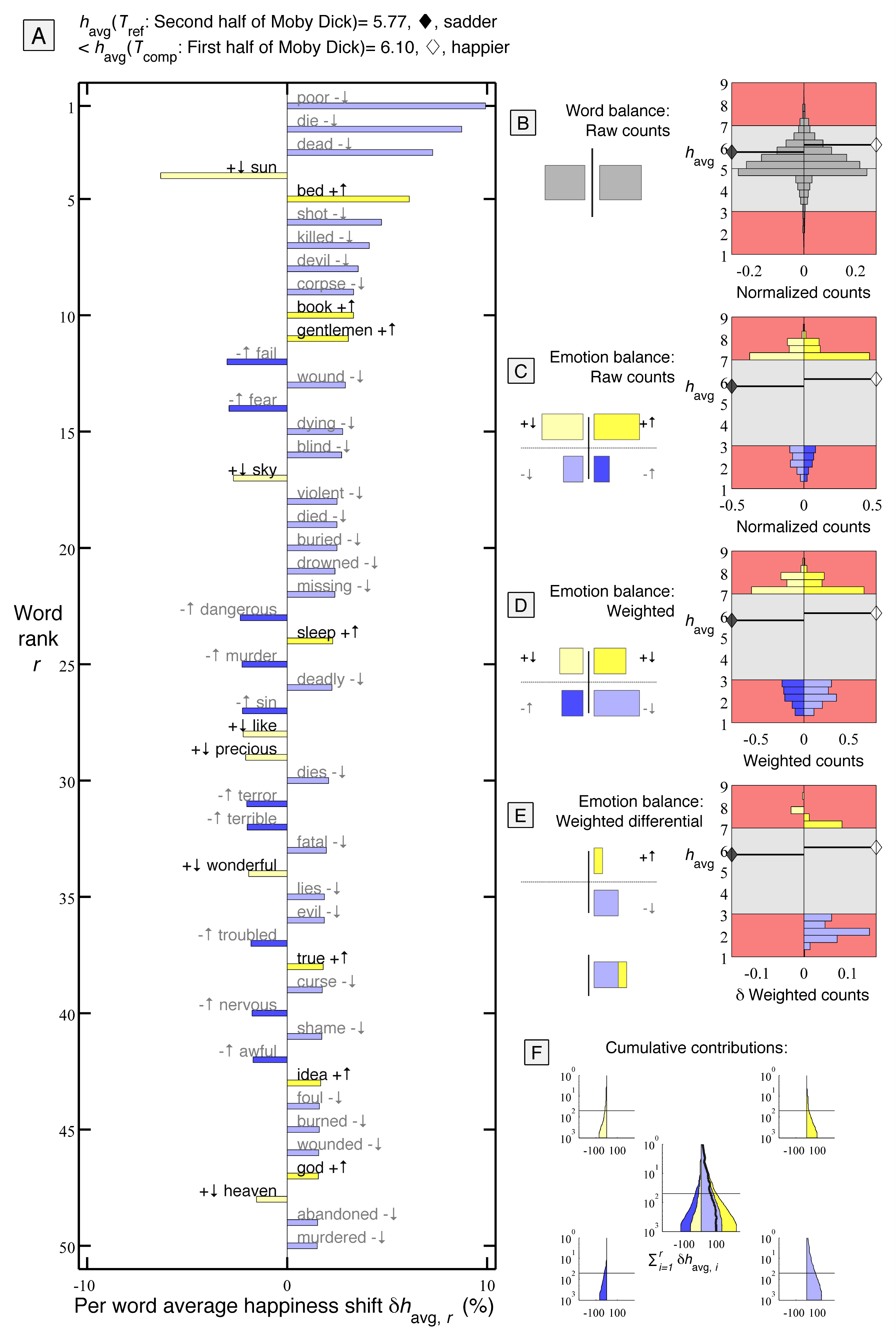}
  \caption{
    Detailed version of the 
    first word shift 
    for Moby Dick
    in Fig.~\ref{fig:mlhap.measurementexamples}.
    See pp.~\pageref{page:mhl.startwordshiftdescription}--\pageref{page:mhl.endwordshiftdescription} for a full explanation.
  }
  \label{fig:universal_wordshift500_figuniversal_wordshift_hli_moby_dick001}
\end{figure*}

\begin{figure*}[tp!]
  \centering
\includegraphics[height=0.9\textheight]{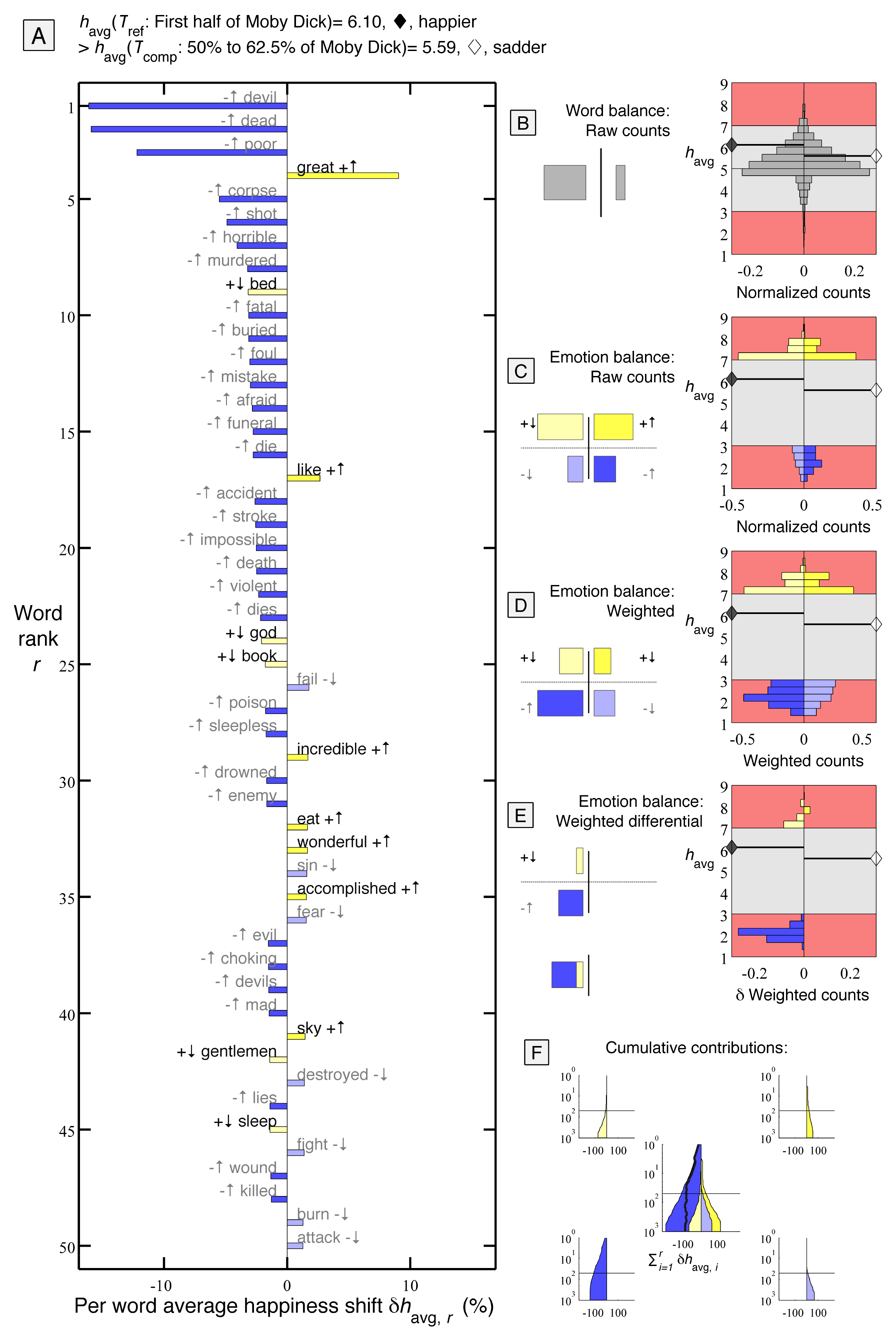}
  \caption{
    Detailed version of the 
    second word shift 
    for Moby Dick
    in Fig.~\ref{fig:mlhap.measurementexamples}.
    See pp.~\pageref{page:mhl.startwordshiftdescription}--\pageref{page:mhl.endwordshiftdescription} for a full explanation.
  }
  \label{fig:universal_wordshift500_figuniversal_wordshift_hli_moby_dick002}
\end{figure*}

\begin{figure*}[tp!]
  \centering
  \includegraphics[height=0.9\textheight]{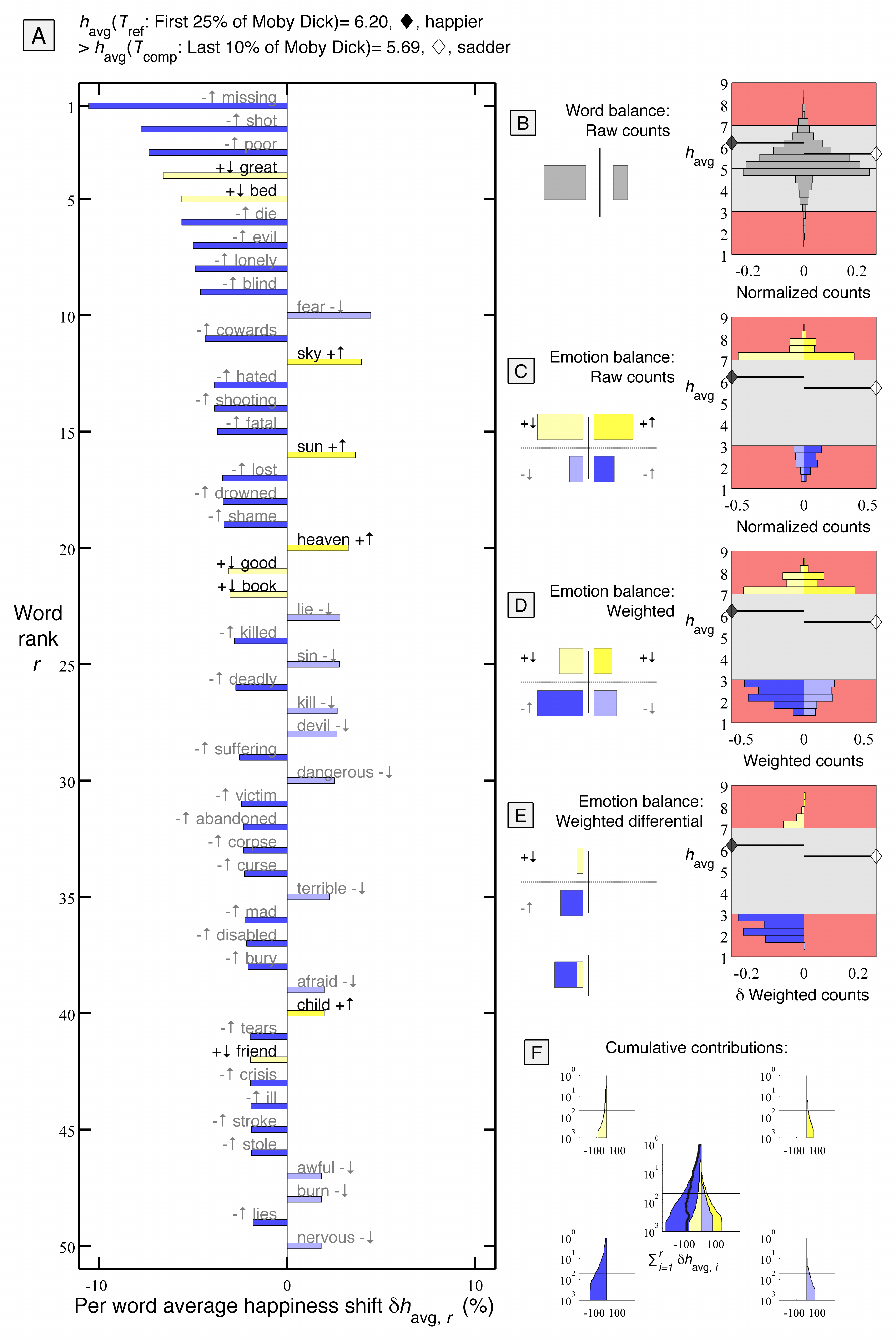}
  \caption{
    Detailed version of the 
    third word shift 
    for Moby Dick
    in Fig.~\ref{fig:mlhap.measurementexamples}.
    See pp.~\pageref{page:mhl.startwordshiftdescription}--\pageref{page:mhl.endwordshiftdescription} for a full explanation.
   }
  \label{fig:universal_wordshift500_figuniversal_wordshift_hli_moby_dick003}
\end{figure*}

\begin{figure*}[tp!]
  \centering
  \includegraphics[height=0.9\textheight]{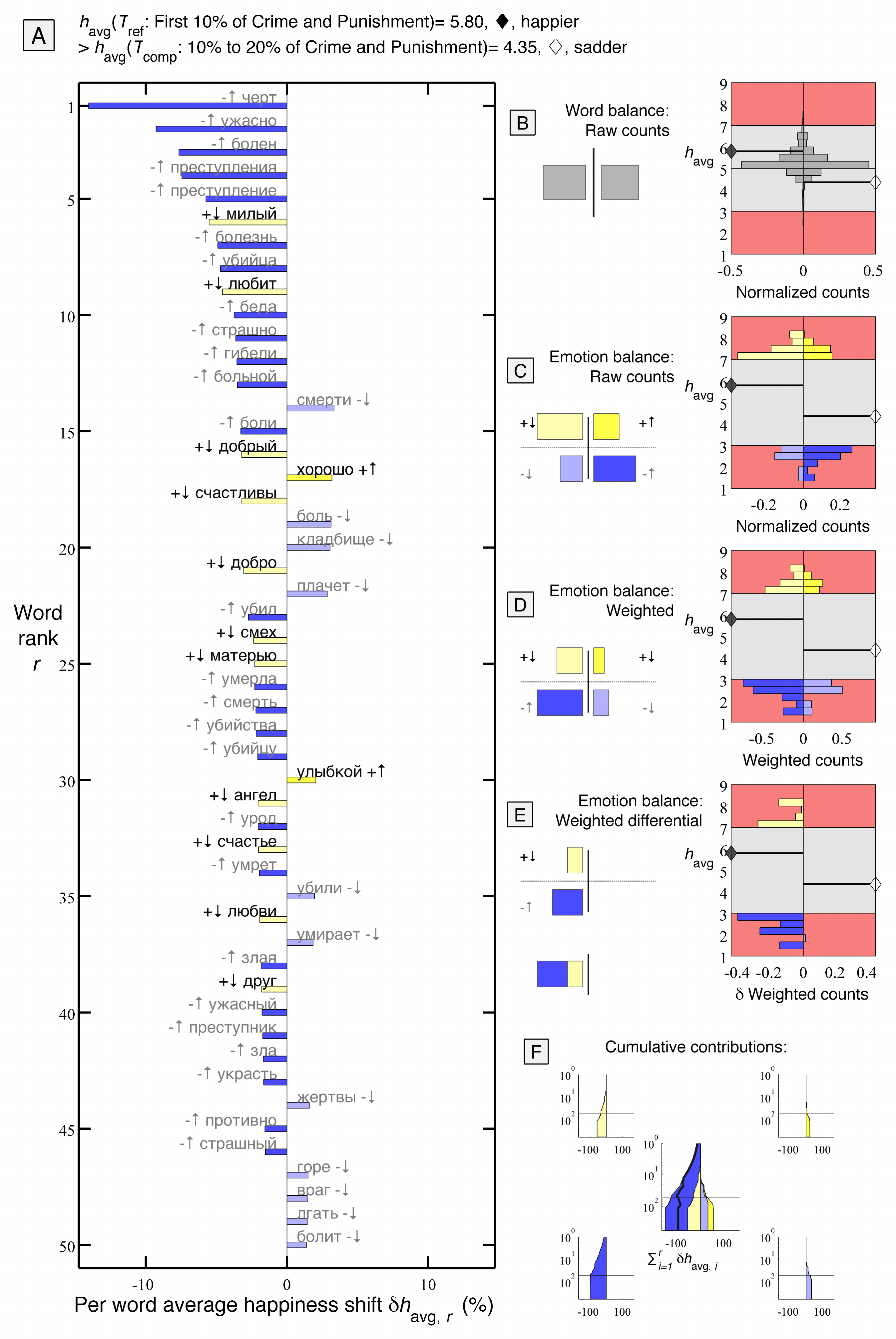}
  \caption{
    Detailed version of the 
    first word shift 
    for Crime and Punishment
    in Fig.~\ref{fig:mlhap.measurementexamples}.
    See pp.~\pageref{page:mhl.startwordshiftdescription}--\pageref{page:mhl.endwordshiftdescription} for a full explanation.
   }
  \label{fig:universal_wordshift500_figuniversal_wordshift_hli_crime_and_punishment001}
\end{figure*}

\begin{figure*}[tp!]
  \centering
  \includegraphics[height=0.9\textheight]{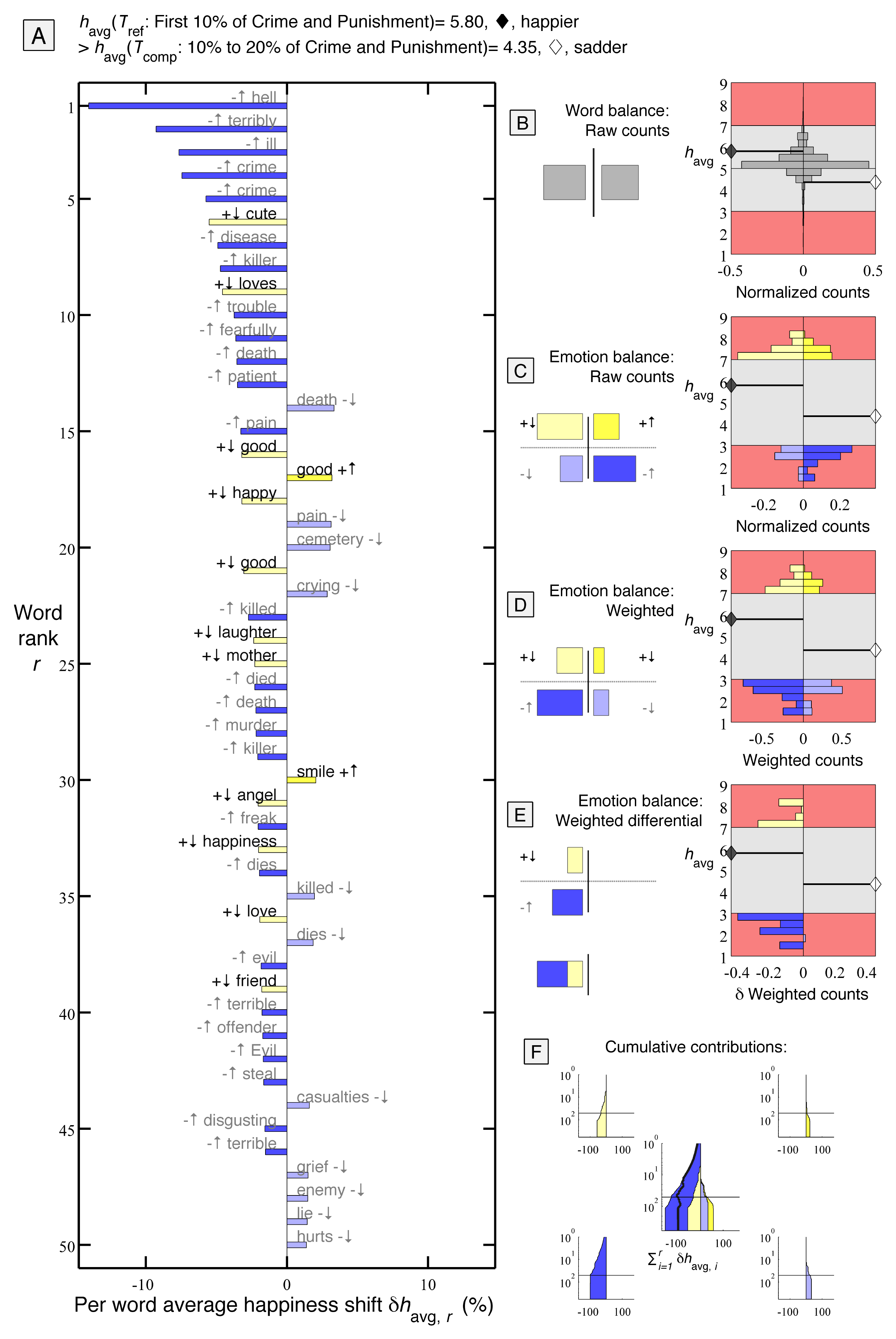}
  \caption{
    Detailed English translation version of the 
    first word shift 
    for Crime and Punishment
    in Fig.~\ref{fig:mlhap.measurementexamples}.
    See pp.~\pageref{page:mhl.startwordshiftdescription}--\pageref{page:mhl.endwordshiftdescription} for a full explanation.
   }
  \label{fig:universal_wordshift500_figuniversal_wordshift_hli_crime_and_punishment001_eng}
\end{figure*}

\begin{figure*}[tp!]
  \centering
  \includegraphics[height=0.9\textheight]{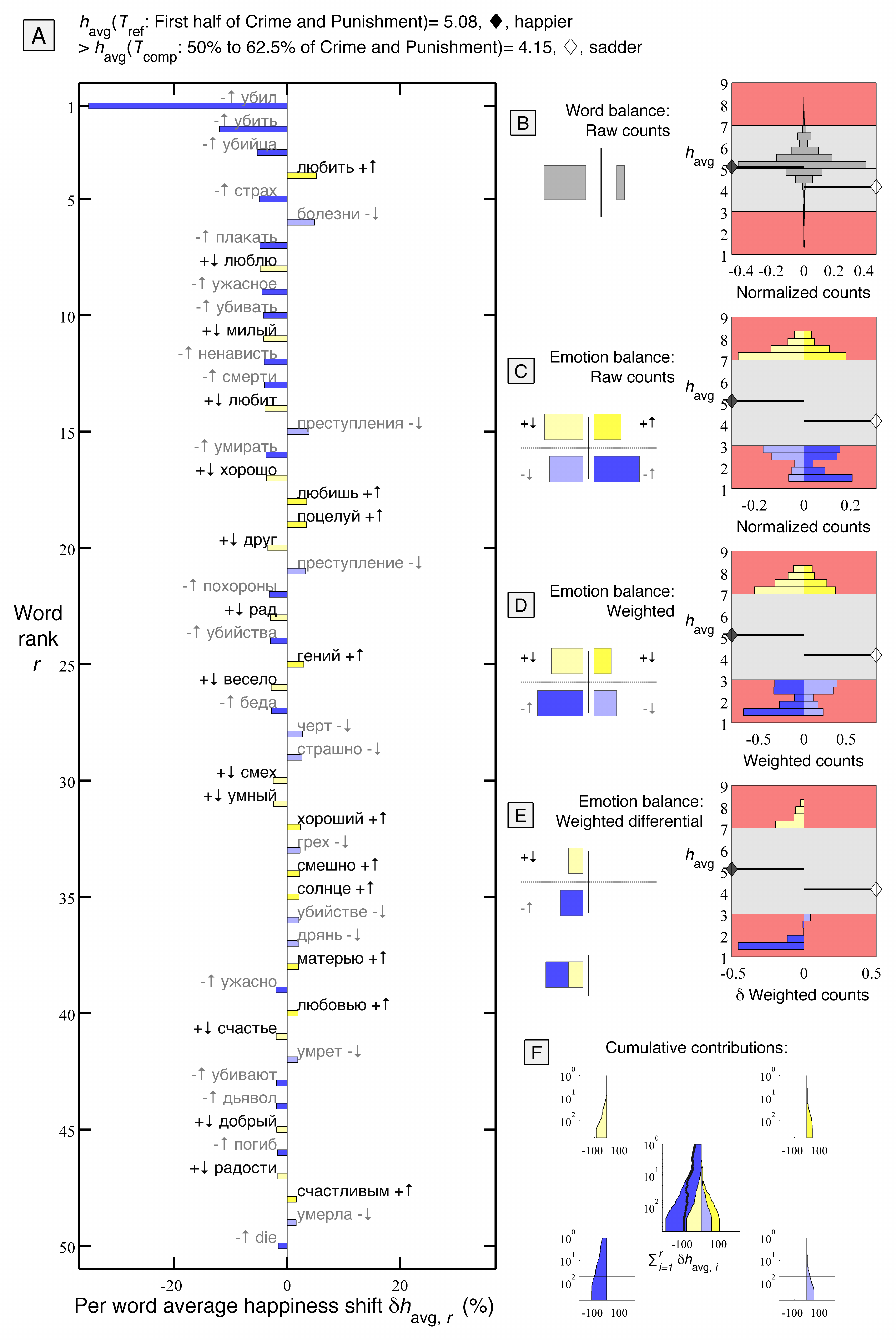}
  \caption{
    Detailed version of the 
    second word shift 
    for Crime and Punishment
    in Fig.~\ref{fig:mlhap.measurementexamples}.
    See pp.~\pageref{page:mhl.startwordshiftdescription}--\pageref{page:mhl.endwordshiftdescription} for a full explanation.
   }
  \label{fig:universal_wordshift500_figuniversal_wordshift_hli_crime_and_punishment002}
\end{figure*}

\begin{figure*}[tp!]
  \centering
  \includegraphics[height=0.9\textheight]{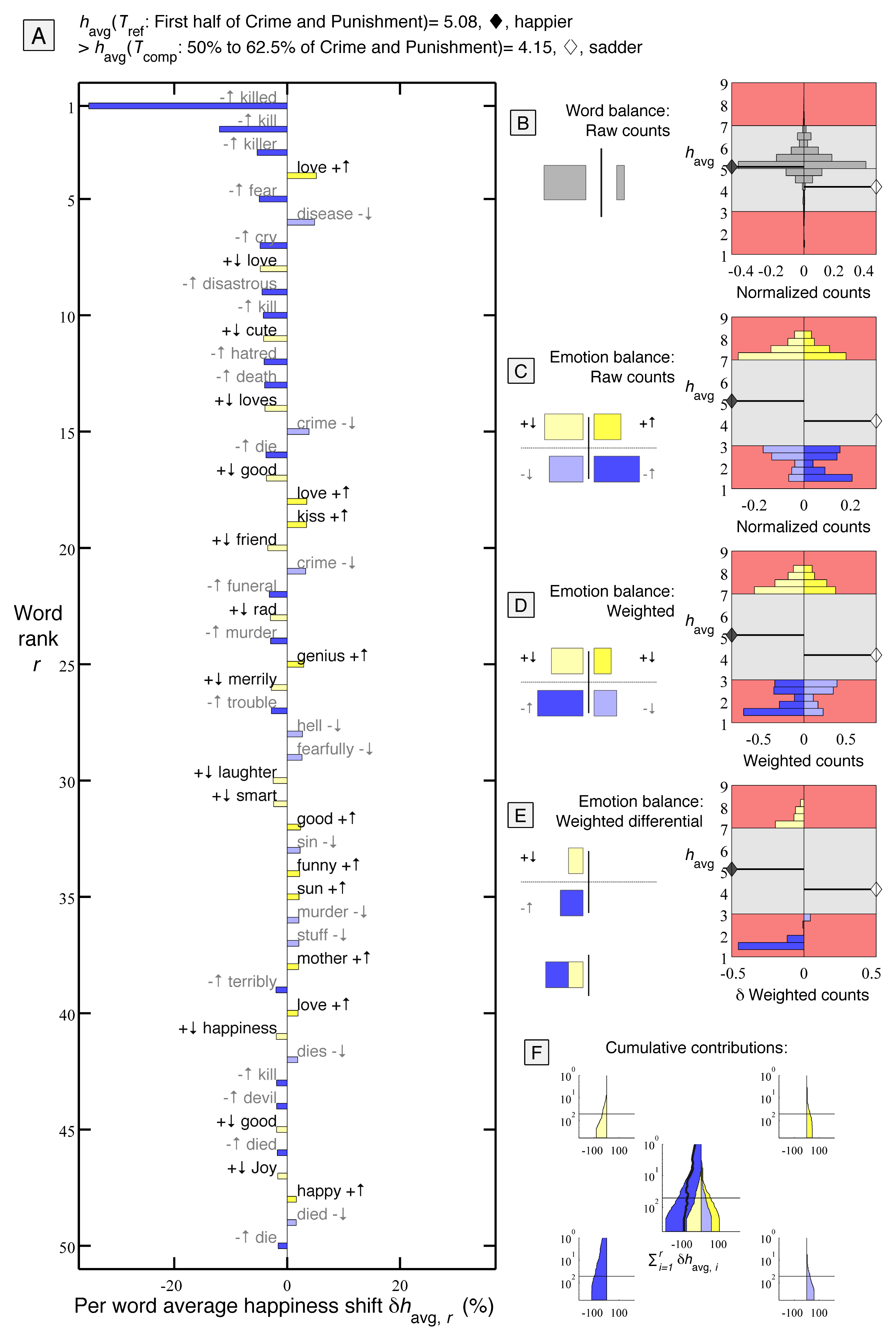}
  \caption{
    Detailed English translation version of the 
    second word shift 
    for Crime and Punishment
    in Fig.~\ref{fig:mlhap.measurementexamples}.
    See pp.~\pageref{page:mhl.startwordshiftdescription}--\pageref{page:mhl.endwordshiftdescription} for a full explanation.
   }
  \label{fig:universal_wordshift500_figuniversal_wordshift_hli_crime_and_punishment002_eng}
\end{figure*}

\begin{figure*}[tp!]
  \centering
  \includegraphics[height=0.9\textheight]{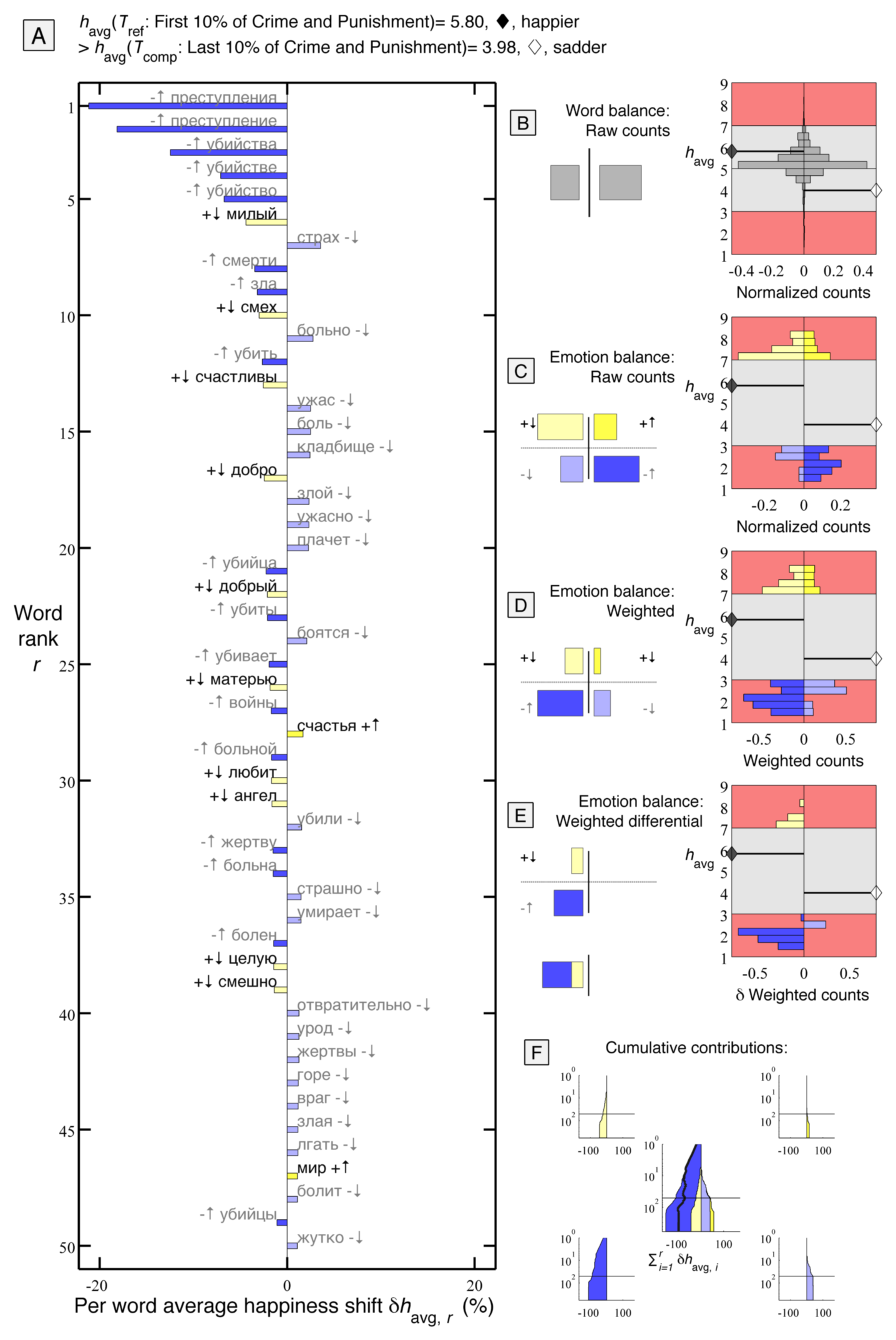}
  \caption{
    Detailed version of the 
    third word shift 
    for Crime and Punishment
    in Fig.~\ref{fig:mlhap.measurementexamples}.
    See pp.~\pageref{page:mhl.startwordshiftdescription}--\pageref{page:mhl.endwordshiftdescription} for a full explanation.
   }
  \label{fig:universal_wordshift500_figuniversal_wordshift_hli_crime_and_punishment003}
\end{figure*}

\begin{figure*}[tp!]
  \centering
  \includegraphics[height=0.9\textheight]{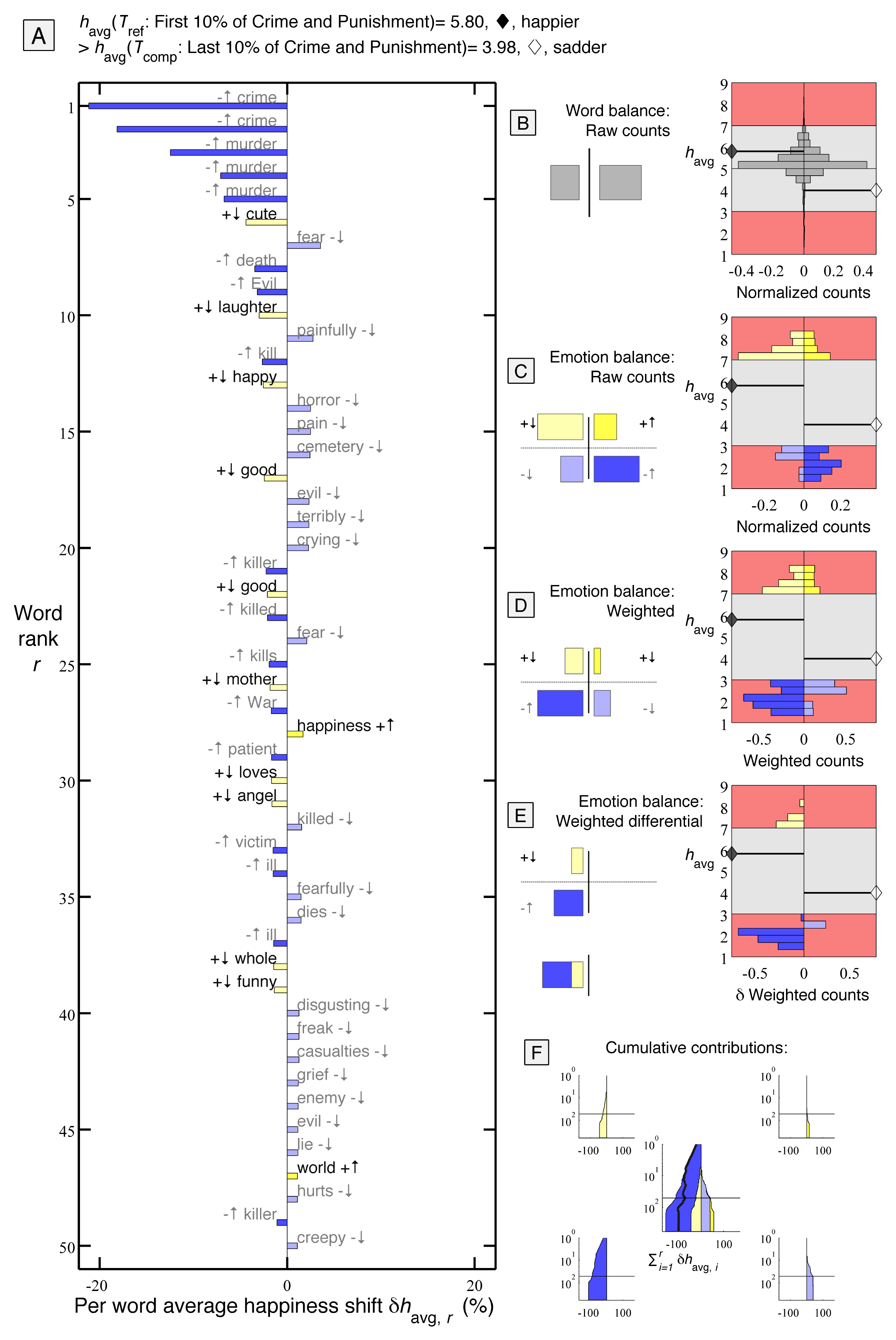}
  \caption{
    Detailed English translation version of the 
    third word shift 
    for Crime and Punishment
    in Fig.~\ref{fig:mlhap.measurementexamples}.
    See pp.~\pageref{page:mhl.startwordshiftdescription}--\pageref{page:mhl.endwordshiftdescription} for a full explanation.
   }
  \label{fig:universal_wordshift500_figuniversal_wordshift_hli_crime_and_punishment003_eng}
\end{figure*}

\begin{figure*}[tp!]
  \centering
  \includegraphics[height=0.9\textheight]{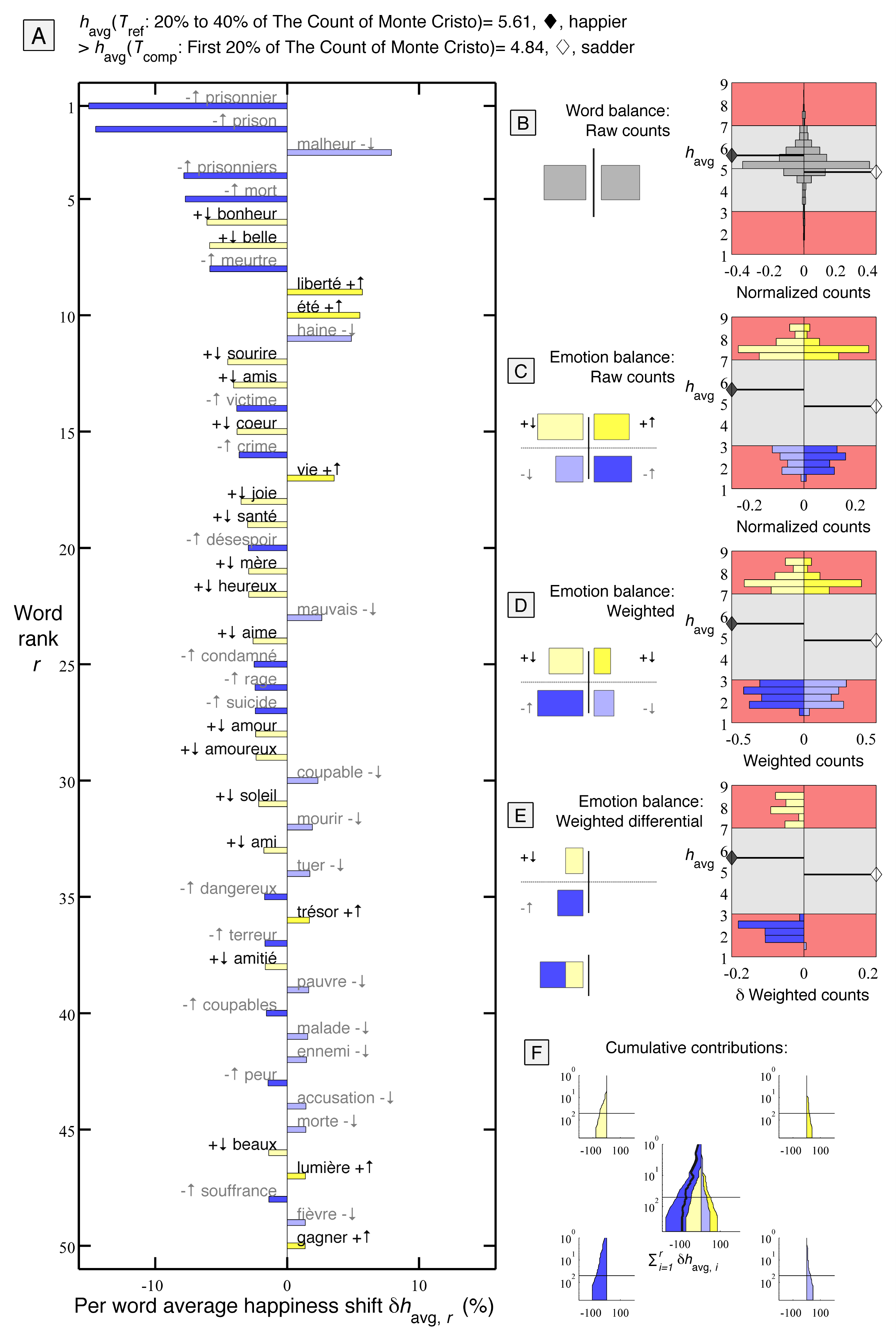}
  \caption{
    Detailed version of the 
    first word shift 
    for the Count of Monte Cristo
    in Fig.~\ref{fig:mlhap.measurementexamples}.
    See pp.~\pageref{page:mhl.startwordshiftdescription}--\pageref{page:mhl.endwordshiftdescription} for a full explanation.
   }
  \label{fig:universal_wordshift500_figuniversal_wordshift_hli_count_of_monte_cristo001}
\end{figure*}

\begin{figure*}[tp!]
  \centering
  \includegraphics[height=0.9\textheight]{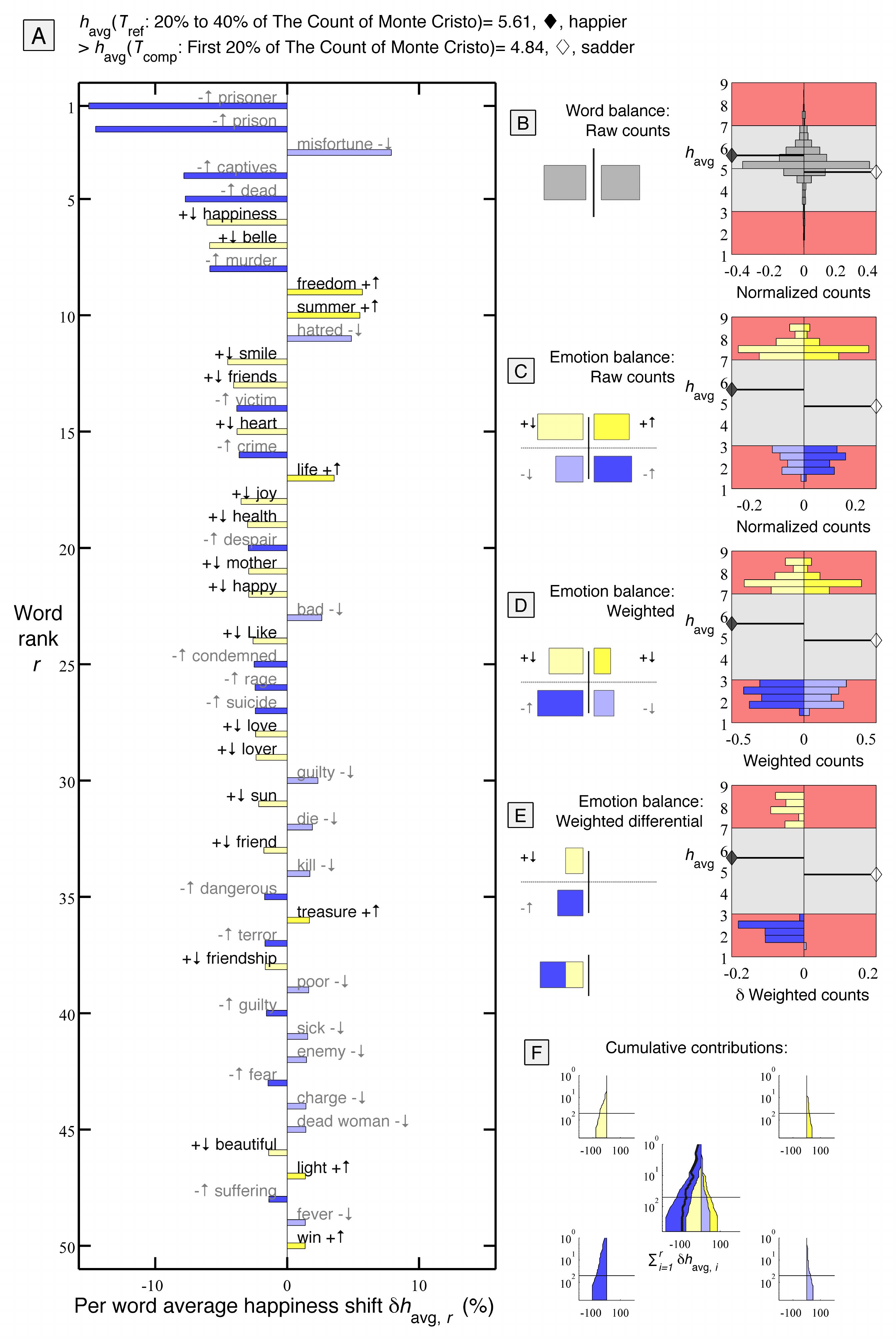}
  \caption{
    Detailed English translation version of the 
    first word shift 
    for the Count of Monte Cristo
    in Fig.~\ref{fig:mlhap.measurementexamples}.
    See pp.~\pageref{page:mhl.startwordshiftdescription}--\pageref{page:mhl.endwordshiftdescription} for a full explanation.
   }
  \label{fig:universal_wordshift500_figuniversal_wordshift_hli_count_of_monte_cristo001_eng}
\end{figure*}

\begin{figure*}[tp!]
  \centering
  \includegraphics[height=0.9\textheight]{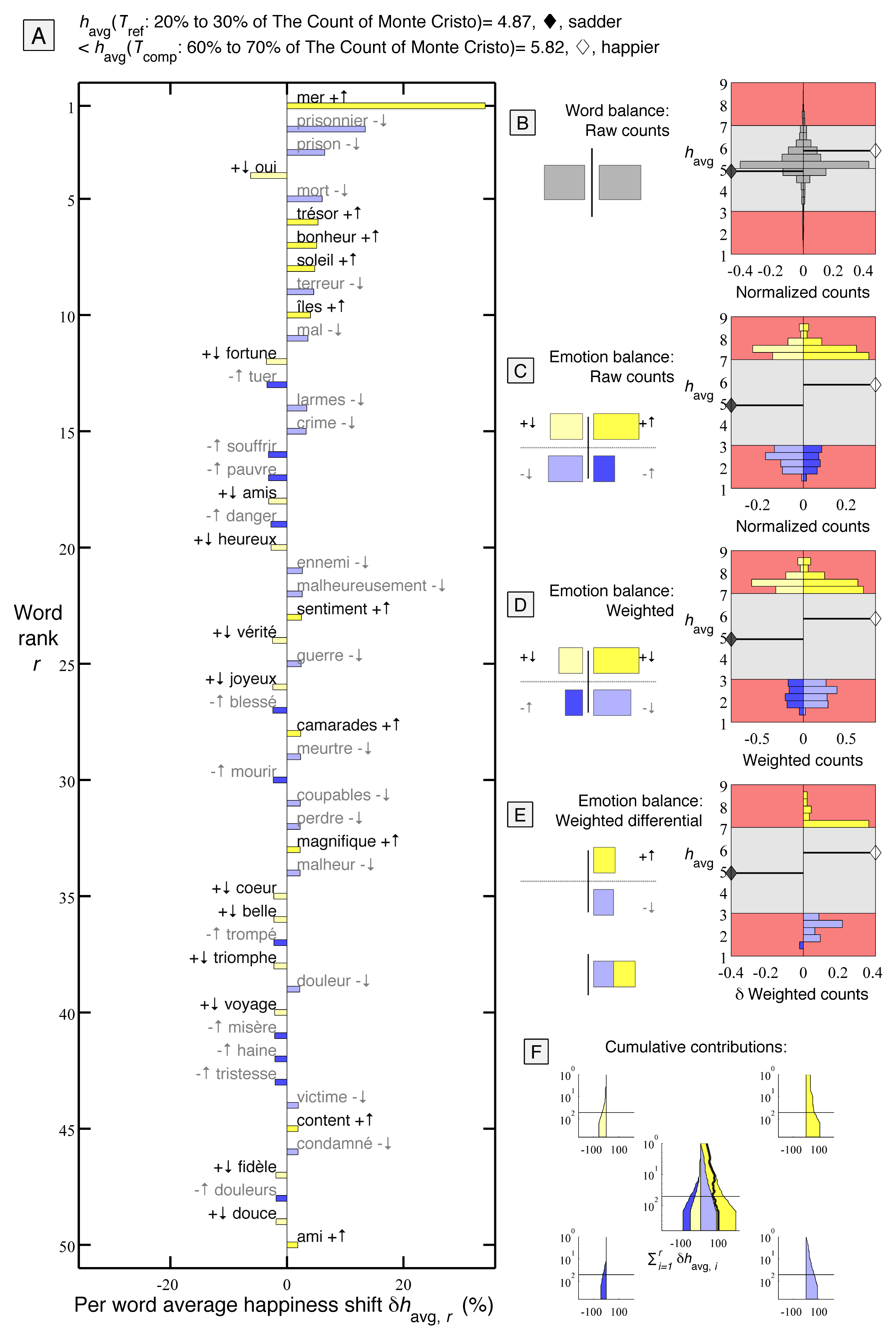}
  \caption{
    Detailed version of the 
    second word shift 
    for the Count of Monte Cristo
    in Fig.~\ref{fig:mlhap.measurementexamples}.
    See pp.~\pageref{page:mhl.startwordshiftdescription}--\pageref{page:mhl.endwordshiftdescription} for a full explanation.
   }
  \label{fig:universal_wordshift500_figuniversal_wordshift_hli_count_of_monte_cristo002}
\end{figure*}

\begin{figure*}[tp!]
  \centering
  \includegraphics[height=0.9\textheight]{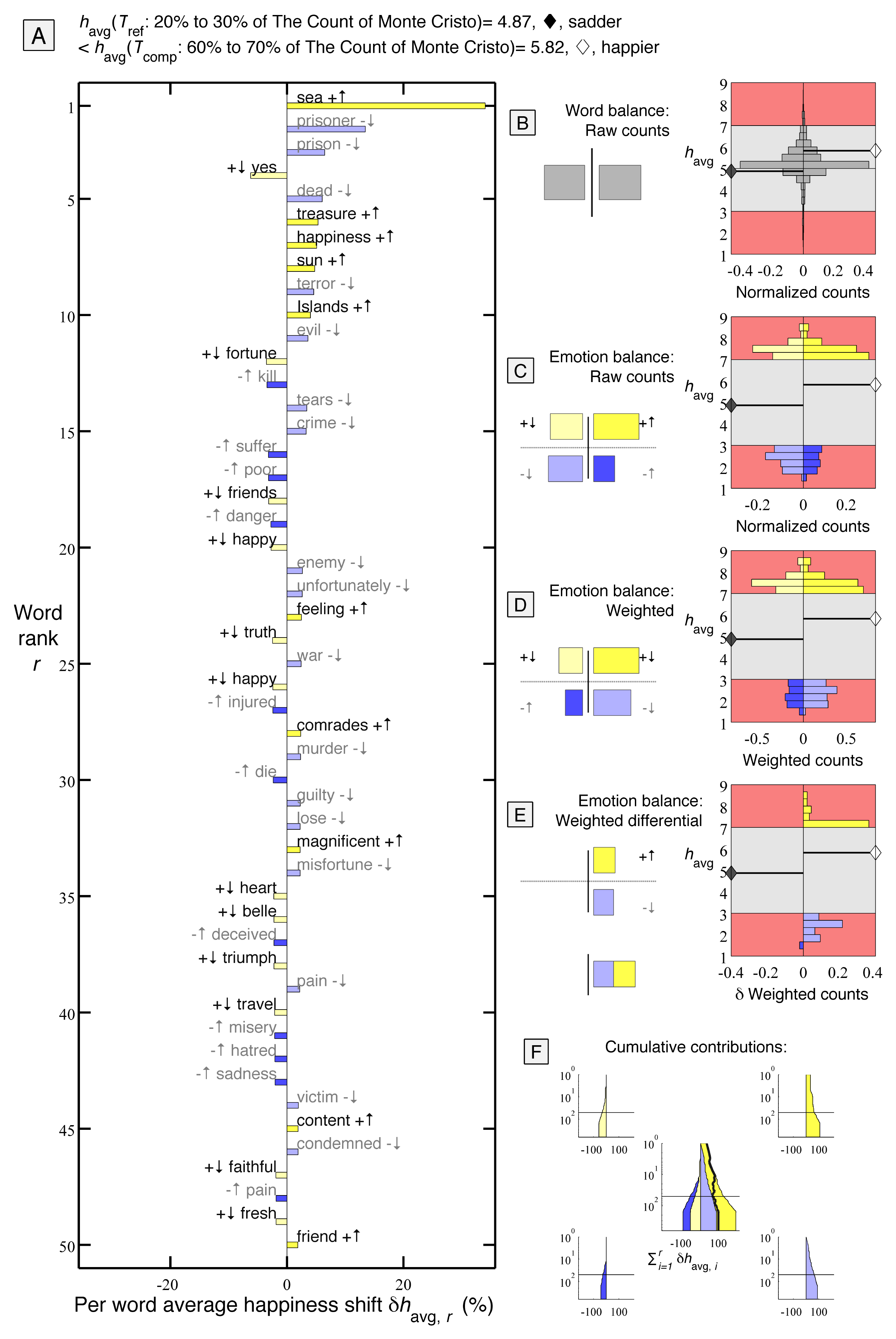}
  \caption{
    Detailed English translation version of the 
    second word shift 
    for the Count of Monte Cristo
    in Fig.~\ref{fig:mlhap.measurementexamples}.
    See pp.~\pageref{page:mhl.startwordshiftdescription}--\pageref{page:mhl.endwordshiftdescription} for a full explanation.
   }
  \label{fig:universal_wordshift500_figuniversal_wordshift_hli_count_of_monte_cristo002_eng}
\end{figure*}

\begin{figure*}[tp!]
  \centering
  \includegraphics[height=0.9\textheight]{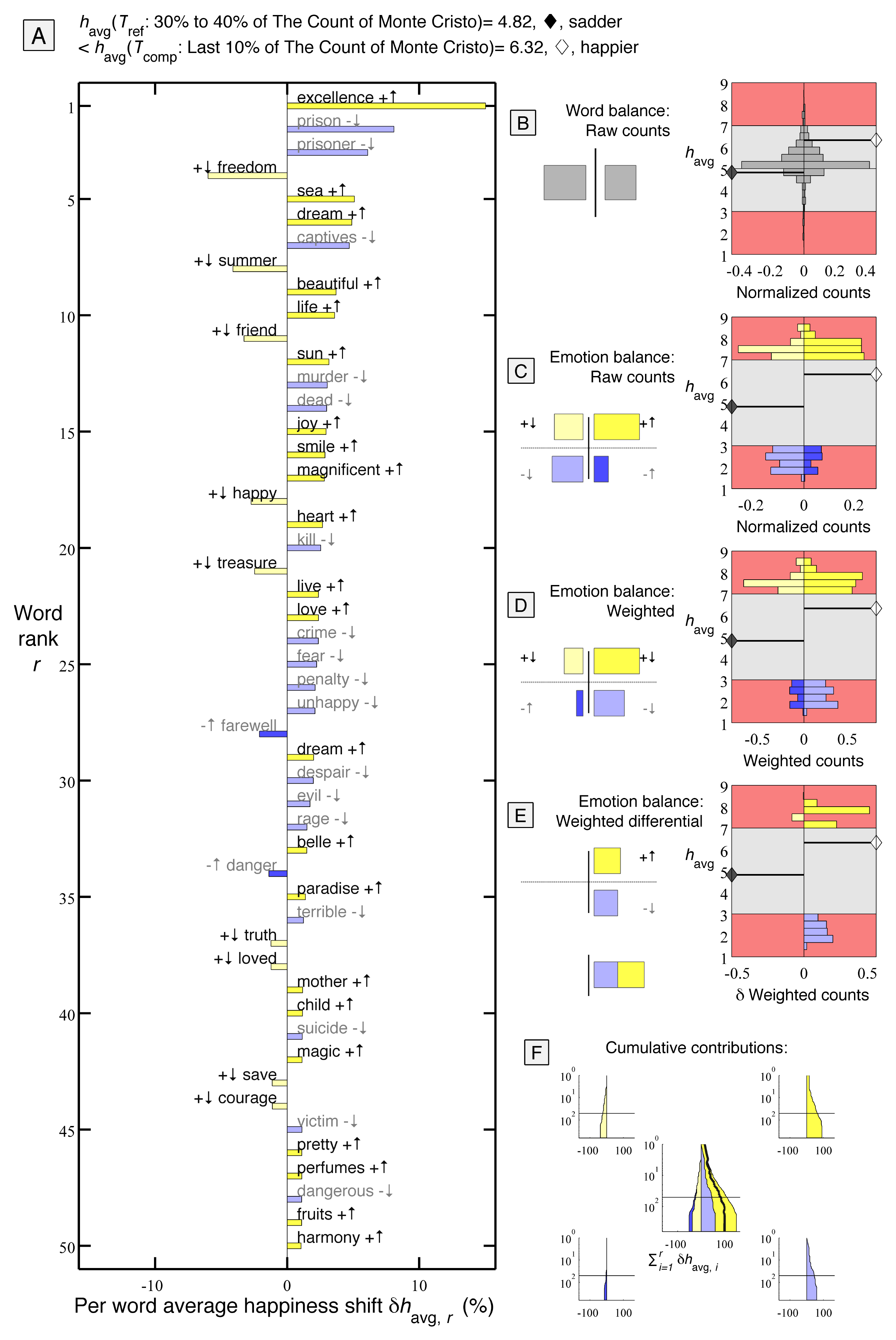}
  \caption{
    Detailed English translation version of the 
    third word shift 
    for the Count of Monte Cristo
    in Fig.~\ref{fig:mlhap.measurementexamples}.
    See pp.~\pageref{page:mhl.startwordshiftdescription}--\pageref{page:mhl.endwordshiftdescription} for a full explanation.
   }
  \label{fig:universal_wordshift500_figuniversal_wordshift_hli_count_of_monte_cristo003_eng}
\end{figure*}

\end{document}